\documentclass[conference]{IEEEtran}
\IEEEoverridecommandlockouts

\usepackage{cite}
\usepackage{amsmath,amssymb,amsfonts}
\usepackage{graphicx}
\usepackage{textcomp}
\usepackage{xcolor}
\def\BibTeX{{\rm B\kern-.05em{\sc i\kern-.025em b}\kern-.08em
    T\kern-.1667em\lower.7ex\hbox{E}\kern-.125emX}}

\usepackage{kotex}
\usepackage{tabularx}
\usepackage{multirow}
\usepackage{url}

\usepackage{algorithm} 
\usepackage{algpseudocode}
\usepackage{graphicx}

\usepackage{float}
\usepackage{comment}
\usepackage[normalem]{ulem}

\usepackage[most]{tcolorbox}
\usepackage{tikz}

\usepackage{booktabs}

\usepackage{caption}
\usepackage{subcaption}
\usepackage{color}
\usepackage{pifont}
\usepackage{setspace}
\usepackage[colorlinks=true,linkcolor=black,citecolor=black,urlcolor=black]{hyperref}
\usepackage{paralist}
\newcommand{\tbdfixed}{{{{\textsc{Oasis}}}}}
\newcommand{\frontend}{{{{\textsc{Oasis-FE}}}}}
\newcommand{\arraynode}{{{{\textsc{Oasis-A}}}}}
\newcommand{\vqoa}{{{{\text{SODA}}}}}

\newcommand{\cb}{\textcolor{blue}}
\newcommand{\cred}{\textcolor{red}}

\newcommand{\bfit}[1]{\textbf{\textit{#1}}}
\newcommand{\squishlist}{
\begin{list}{$\bullet$}
	{ \setlength{\itemsep}{0pt}      \setlength{\parsep}{-0pt}
		\setlength{\topsep}{4pt}       \setlength{\partopsep}{0pt}
		\setlength{\listparindent}{-2pt}
		\setlength{\itemindent}{-5pt}
		\setlength{\leftmargin}{1em} \setlength{\labelwidth}{0em}
		\setlength{\labelsep}{0.5em} } }
\newcommand{\squishend}{
\end{list}  }

\usepackage{circledsteps}
\newcommand{\CircledTextBlack}[2][]{%
    \CircledParamOpts{inner color=white, outer color=none, fill color=black, inner xsep=2pt, inner ysep=2pt, #1}{1}{#2}%
}

\title{\huge\tbdfixed{}:
Object-based Analytics Storage for Intelligent SQL Query Offloading in Scientific Tabular Workloads}

\author{
Soon Hwang$^1$, {\href{https://orcid.org/0009-0008-9293-173X}{Junhyeok Park}}$^1$, Junghyun Ryu$^1$, Seonghoon Ahn$^1$, Jeoungahn Park$^2$, Jeongjin Lee$^2$\\
Soonyeal Yang$^2$, Jungki Noh$^2$, Woosuk Chung$^2$, Hoshik Kim$^2$, and Youngjae Kim$^1$\\
$^1$Sogang University, Seoul, Republic of Korea \quad $^2$Memory Systems Research, SK hynix Inc.\thanks{Y. Kim is the corresponding author.}
}

\begin{document}

\maketitle

\begin{abstract}
Computation-Enabled Object Storage (COS) systems, such as MinIO and Ceph, have recently emerged as promising storage solutions for post hoc, SQL-based analysis on large-scale datasets in High-Performance Computing (HPC) environments.
By supporting object-granular layouts, COS facilitates column-oriented access and supports in-storage execution of data reduction operators, such as filters, close to where the data resides.
Despite growing interest and adoption, existing COS systems exhibit several fundamental limitations that hinder their effectiveness.
First, they impose rigid constraints on output data formats, limiting flexibility and interoperability.
Second, they support offloading for only a narrow set of operators and expressions, restricting their applicability to more complex analytical tasks.
Third--and perhaps most critically--they fail to incorporate design strategies that enable compute offloading optimized for the characteristics of deep storage hierarchies.
To address these challenges, this paper proposes \tbdfixed{}, a novel COS system that features:
(i) flexible and interoperable output delivery through diverse formats, including columnar layouts such as Arrow;
(ii) broad support for complex operators (e.g., aggregate, sort) and array-aware expressions, including element-wise predicates over array structures; and
(iii) dynamic selection of optimal execution paths across internal storage layers, guided by operator characteristics and data movement costs.
We implemented a prototype of \tbdfixed{} and integrated it into the Spark analytics framework.
Through extensive evaluation using real-world scientific queries from HPC workflows, \tbdfixed{} achieves up to a 32.7\% performance improvement over Spark configured with existing COS-based storage systems.
\end{abstract}

\section{Introduction}
\label{sec:intro}

In modern scientific research, vast amounts of data are generated through simulations and experiments in domains such as Computational Fluid Dynamics (CFD), High-Energy Physics (HEP), and Particle-In-Cell (PIC) simulations.
These datasets are typically structured in tabular formats with well-defined schemas, where each record--such as a CFD cell or particle--contains a consistent set of fields~\cite{HEPtabular, gray2005scientific, particletabular_liu2022processing, CFPtabular_da2011framework}.
This tabular structure aligns well with the relational model, enabling researchers to perform post hoc analysis using SQL-style queries (§\ref{sec:back_post_hoc})\cite{roi_chiu2015memory}.
To support scalable analysis of such large datasets, distributed data processing frameworks--most notably Apache Spark\cite{spark}--have been widely adopted on High-Performance Computing (HPC) systems for scientific data analytics~\cite{2014parallelstructured, Gu_Springer_2018}.

Meanwhile, recent advances in HPC simulations and scientific instrumentation have further accelerated data growth, placing increasing pressure on analytics systems to keep pace~\cite{grider2024snia}.
For example, supercomputers such as Frontier\cite{frontier} at Oak Ridge National Laboratory (ORNL) enable massive simulations that produce structured outputs across thousands of timesteps.
Likewise, facilities like the High-Luminosity Large Hadron Collider (HL-LHC)\cite{hl-lhc-cern} substantially increase data acquisition rates through enhanced detector resolution and higher event frequency.
As data volumes continue to grow, their rate of increase is rapidly outpacing improvements in I/O and network bandwidth, leading to two key challenges: 

First, data movement has become a major bottleneck in HPC analytics, exacerbated not only by increasing data volumes but also by low-selectivity queries that focus on narrow regions of interest~(§\ref{sec:low_selectivity}).
While processing power has advanced for large-scale post-hoc analysis, storage I/O has not kept pace, causing transfer delays, idle compute nodes, and reduced system efficiency~\cite{xing2018arraybridge, NERSC_datamovement, liang2023survey, Cicotti2016}.
Second, traditional POSIX-based file systems struggle with efficient data placement at scale~\cite{IHEP-object}. Their flat byte-stream model lacks structural awareness, making it difficult to selectively place hot columns on fast storage. 
This often results in hot data being placed on slow storage or in fast-tier overuse, degrading I/O performance~\cite{smith2023cephs3objectdata}.
{These limitations highlight the need for storage systems that minimize data movement and support column-aware layout} (§\ref{sec:back_challenge}).

To address these challenges, Computation-Enabled Object Storage (COS) systems--such as MinIO~\cite{minio}, Ceph~\cite{ceph} with S3 Select~\cite{S3select}, and Ceph with SkyhookDM support~\cite{skyhookdm_paper}--have emerged as promising solutions that augment object storage with lightweight computation at the storage layer (§\ref{sec:back_cos})~\cite{minio-mc-sql, ceph-s3-select, skyhookdm_paper, sim2015analyzethis, sim2017tagit}.
These systems can perform \textit{filter} and \textit{project} operations near the data, mitigating data movement bottlenecks in HPC environments.
COS also improves data placement by supporting column-level granularity through its object abstraction, enabling selective tiering based on access frequency. This addresses the data placement limitations of POSIX systems and aligns data layout with workload behavior.

To enable compute offloading to storage, COS systems are designed following two distinct architectural approaches: executing query operations at the interface layer (e.g., Ceph S3 Select~\cite{ceph-s3-select}, MinIO Select~\cite{minio-mc-sql}), or embedding computation within the internal layers of the storage stack (e.g., SkyhookDM~\cite{skyhookdm_paper}).
While these systems differ in their offloading mechanisms, they share common limitations that hinder support for complex, high-throughput analytics. These limitations stem from architectural constraints and limited expressiveness in query execution, and are summarized as follows (§\ref{subsec:limit}).

\squishlist
\item 
\textbf{Inflexible Output Format Limits Optimization and Integration:} 
Existing COS systems return offloaded query results in fixed output formats such as CSV and JSON, which lack structural metadata and require additional parsing, or Apache Arrow~\cite{Arrow}, which offers efficient columnar access but is not universally supported by all analytics engines.
\item \textbf{Limited Offloading Capability for Various Operators and Array Semantics:} 
{Most existing COS systems support only \textit{filter} and \textit{project} operations with scalar conditions, lacking essential advanced operators such as \textit{aggregate} and array-based expressions common in HPC queries~(§\ref{sec:observation}).} 
As a result, for queries involving such unsupported operators or array-based conditions, COS systems still need to transfer entire files or row groups to the compute layer, increasing data movement and slowing analysis.

\item \textbf{Excessive Inter-Storage Data Movement Due to Fixed Execution Layer:} Current COS systems execute queries at a single fixed layer, such as the gateway node that interfaces with compute clients, and thus fail to minimize internal data transfers. 
This inflexible design prevents early-stage data reduction in lower layers of the storage stack, such as storage array nodes, thus limiting the benefits of offloading.


\squishend

To overcome the aforementioned limitations, we propose \tbdfixed{}, a new COS system designed for analytical workflows in HPC environments (§\ref{sec:design_principle}).
\tbdfixed{} is composed of a frontend node (\frontend{}) as a object interface gateway and multiple storage arrays (\arraynode{}), interconnected via high-speed NVMe-over-Fabrics (NVMe-oF) over RDMA (§\ref{sec:design_overview}).

\tbdfixed{} implements the following key features:
\textbf{First}, \tbdfixed{} generates both intermediate and final results of offloaded query execution in the Arrow columnar format, within storage nodes and during transmission to compute nodes. This design reduces serialization overhead and enables efficient data exchange across system layers.
Final outputs can also be serialized in CSV or JSON format for compatibility with legacy tools.
\textbf{Second}, \tbdfixed{} provides advanced operators such as aggregation, sorting, and array-level computations, including element-wise arithmetic and conditional evaluation. 
These operations are executed directly by the storage-layer Query Executor, which is implemented using DuckDB~\cite{DuckDB} due to its lightweight design, embeddability, and support for complex SQL semantics that align well with \tbdfixed{}’s core requirements (§\ref{subsec:Result_Handler}).
\textbf{Third}, \tbdfixed{} deploys Query Executors on both the \frontend{} and the \arraynode{}.
It then employs a Local Optimizer that analyzes the offloaded query plan and applies the \underline{S}torage-side Query Plan \underline{O}ffloading and \underline{D}ecomposition \underline{A}lgorithm (\vqoa{}) to partition the plan into subplans for distributed execution across the \frontend{} and \arraynode{} nodes (§\ref{subsec:local_opt}).

Based on operator characteristics, \vqoa{} selects between two decomposition strategies (§\ref{subsec:algo}): (1) Coefficient-Aware Decomposition~(CAD) and (2) Structure-Aware Placement~(SAP).
CAD applies to queries with scalar comparisons or simple scalar computations, where data movement can be reasonably estimated using precomputed statistics. It identifies a partitioning point that minimizes total data movement based on estimated input-output ratios.
In contrast, SAP targets queries for which data movement is difficult to estimate by statistics, such as those with array-level conditions or computations. SAP executes these operators close to the data and employs a lazy strategy. Specifically, it measures intermediate result sizes at runtime and forwards results to upper layers only when they fit within available internal bandwidth.

\begin{figure}[!t]
	\centering
	\includegraphics[width=1.0\linewidth]{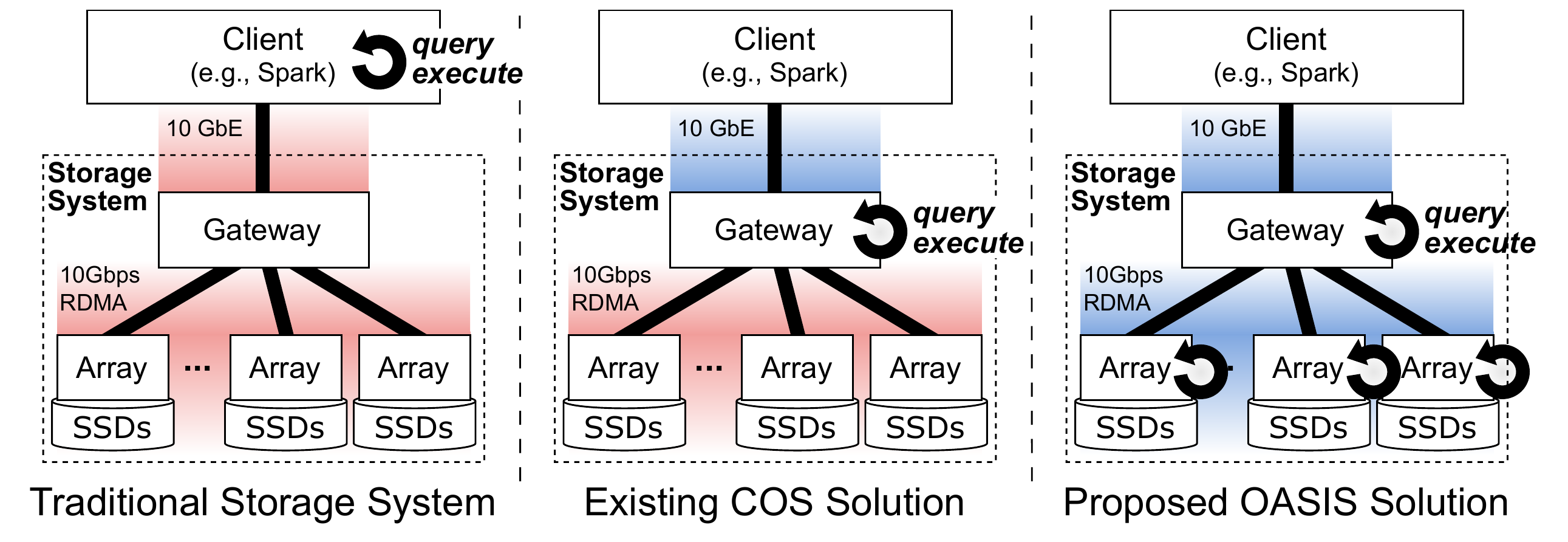}
	\vspace{-15pt}
	\caption{Comparison of traditional storage systems, existing COS solutions, and the \tbdfixed{} system for data analytics. 
    }
	\label{fig:system_compare}
	\vspace{-12pt}
\end{figure}

Figure~\ref{fig:system_compare} compares the storage architecture of \tbdfixed{} with that of existing COS solutions.
While existing COS systems reduce storage-to-compute data transfers by executing queries at the gateway layer within the storage system, they still incur substantial internal data movement across layers.
In contrast, \tbdfixed{} minimizes both inter-layer and storage-to-compute data transfers by initiating early-stage query processing in lower storage layers, guided by data movement cost.

\begin{figure*}[!t]
	\centering
	\includegraphics[width=0.95\linewidth]{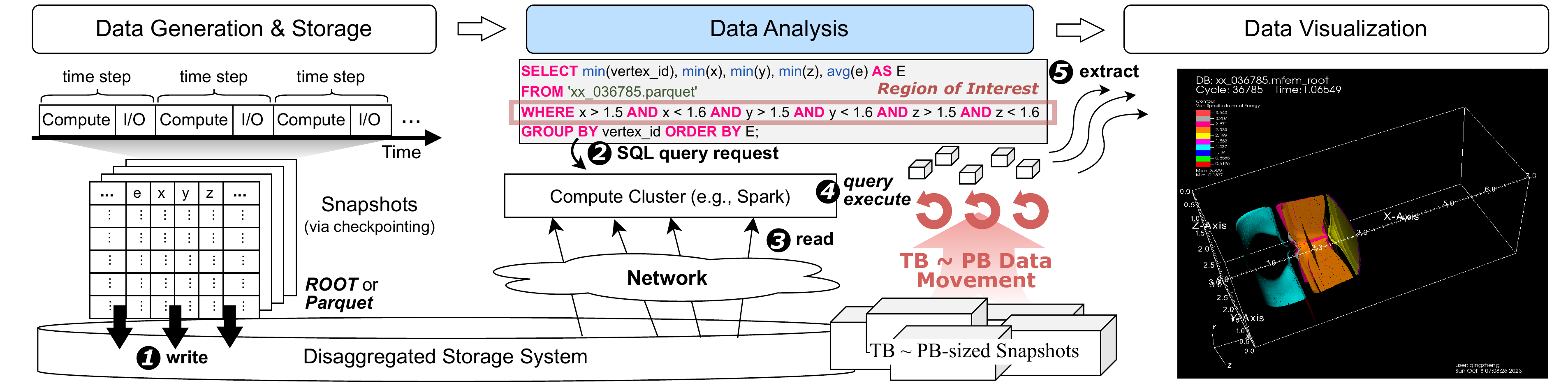}
	\vspace{-2pt}
	\caption{A scientific analytics workflow, illustrating a post-hoc analysis pipeline  (\textit{\CircledTextBlack{1}} - \textit{\CircledTextBlack{5}}) based on the disaggregated storage infrastructure. The example SQL query and visualization results are from the open-source Laghos 3D Mesh Dataset~\cite{laghos_github}.
    }
	\label{fig:back_long}
	\vspace{-8pt}
\end{figure*}


We implemented a prototype of \tbdfixed{} by building key components on SPDK~v23.09 and DuckDB~v1.3.0, where DuckDB serves as the in-storage Query Executor deployed across \frontend{} and \arraynode{}.
We integrated \tbdfixed{} with Apache Spark~4.0.0 to support query offloading. 
Our evaluation was conducted on a multi-node Spark cluster and an RDMA-connected \tbdfixed{} setup, comprising 36–112 core servers with up to 386\,GB of memory.

We evaluated \tbdfixed{} using a diverse set of real-world analytical queries from representative workflows, including CFD and HEP domains.  
Evaluation results show that \tbdfixed{} achieves up to 70.59\% speedup over the traditional storage system and up to 32.7\% speed up compared to existing COS-based setups.
Furthermore, the \vqoa{}  demonstrated its effectiveness in minimizing query execution time by optimally partitioning the query plan for complex queries composed of multiple operators (e.g., Q1 in Table~\ref{tab:queries}).

In summary, our key contributions are as follows:
\squishlist
\item{
We collect and analyze real-world tabular queries from HPC workflows in domains such as CFD, HEP, and PIC, with a focus on identifying operator patterns and structural traits that impact data movement and offloading opportunities.
}
\item{
We design and implement \tbdfixed{}, a novel object-based, computation-enabled analytical storage system that supports columnar formats, advanced operators with array-level expressions, and dataflow-aware query path optimization across hierarchical internal storage layers.
}
\item{
We propose \vqoa{}, a query decomposition mechanism that partitions full offloaded query plans into subplans for hierarchical internal storage layers of \frontend{} and \arraynode{}. It applies CAD for scalar-centric workloads and SAP for array-centric workloads to enable efficient offloading.
}
\squishend

\section{Background and Related Work}
\label{sec:back}

\subsection{Post-Hoc Analysis Workflow in Scientific Applications}
\label{sec:back_post_hoc}



Figure~\ref{fig:back_long} presents a representative post-hoc analysis workflow widely adopted in modern scientific applications, including those in Computational Fluid Dynamics (CFD), High-Energy Physics (HEP), and Particle-In-Cell (PIC) simulations. 
This workflow typically consists of four sequential stages: \textit{data generation}, \textit{data storage}, \textit{data analysis}, and \textit{data visualization}.

\subsubsection{\textbf{Data Generation}}
Modern scientific applications generate data by computing physical quantities (e.g., position, velocity, and energy) of individual particles within defined simulation domains, such as observation regions or grid cells. 
In observation-driven applications, such as particle collision experiments, data is collected in real-time through specialized detectors and instrumentation systems.


\subsubsection{\textbf{Data Storage}}
The generated data is periodically persisted via checkpointing (\textbf{\textit{\CircledTextBlack{1}}}) as schema-consistent records, where identical attributes are recorded per timestep or event.
Hierarchical formats such as HDF5~\cite{sehrish2017spark} and ROOT~\cite{liu2016h5spark} are widely used to organize this data into nested groups and datasets. Due to their repetitive structure, such records are often convertible into tabular form. Recently, columnar formats like Parquet~\cite{Parquet} has gained popularity for analytical workloads, leading to its adoption as a native storage format~\cite{parquetstore_gavalian2025high, parquetstore_planas2018accelerating} and motivating the conversion of HDF5 and ROOT data for improved compatibility and scalability~\cite{canali_spark_hep_2024}.


\subsubsection{\textbf{Data Analysis}}
For analyzing tabular data, distributed data processing engines such as Apache Spark~\cite{spark} and Relational Database Management Systems~(RDBMS) are commonly used.
In Spark-based environments,  users can perform efficient analysis on specific Regions of Interest~(ROI) using either SQL queries or high-level API such as the Spark DataFrame API~\cite{Gu_Springer_2018, dong2014parallel, asaadi2016comparative}.
Upon receiving an analysis request (\textbf{\textit{\CircledTextBlack{2}}}), the system loads relevant data from storage into compute nodes (\textbf{\textit{\CircledTextBlack{3}}}), where queries are executed to extract scientifically meaningful subsets (\textbf{\textit{\CircledTextBlack{4}}}).


\subsubsection{\textbf{Data Visualization}}
These extracted subsets are subsequently leveraged for downstream analytical tasks, including visualizing simulation results (\textbf{\textit{\CircledTextBlack{5}}}), comparing with experimental observations, and validating the accuracy and reliability of the simulation outcomes.

\begin{figure}[!h]
    \centering
    \begin{tabular}{@{}c@{}c@{}c@{}}
    \includegraphics[width=0.333\linewidth]{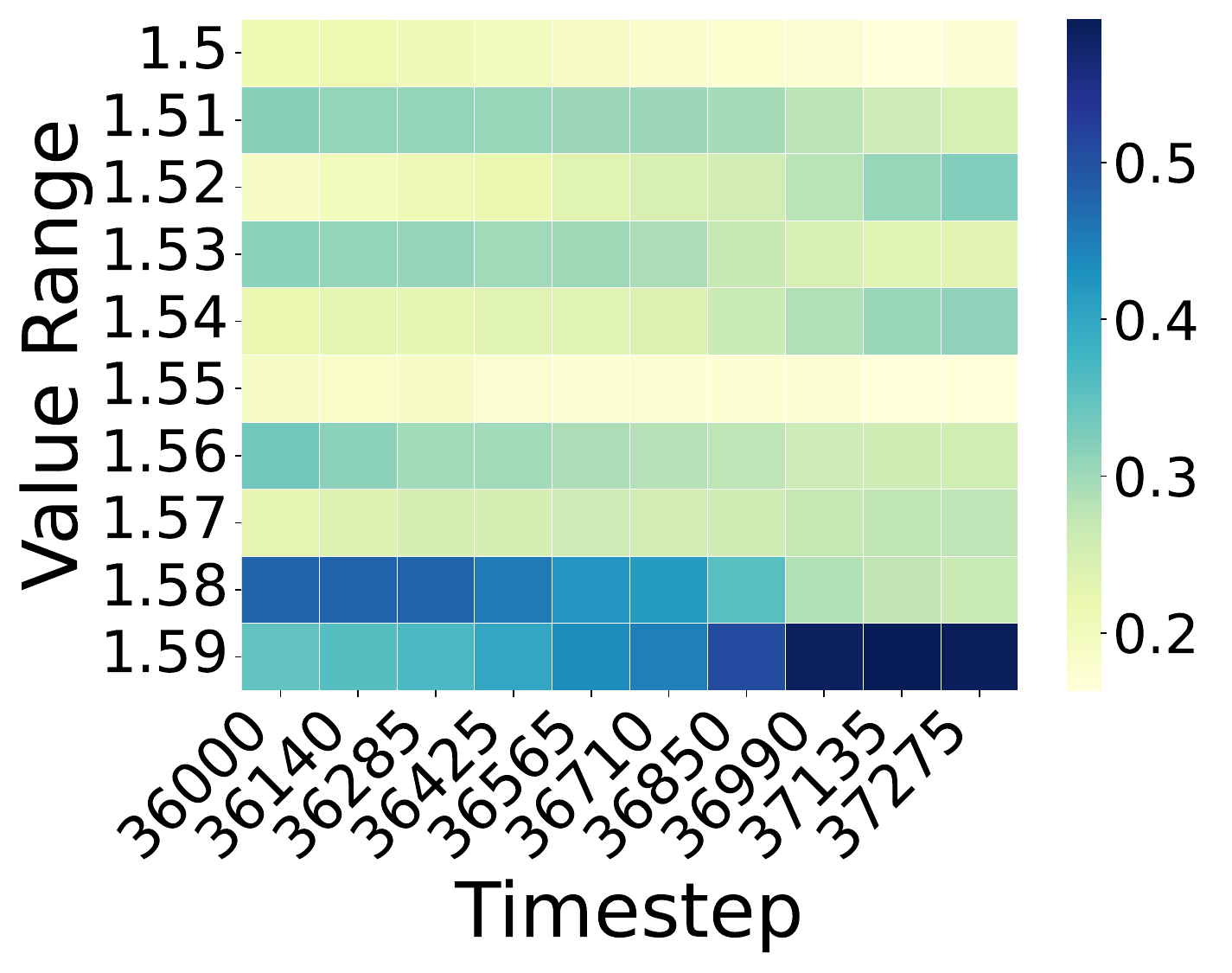} &
    \includegraphics[width=0.333\linewidth]{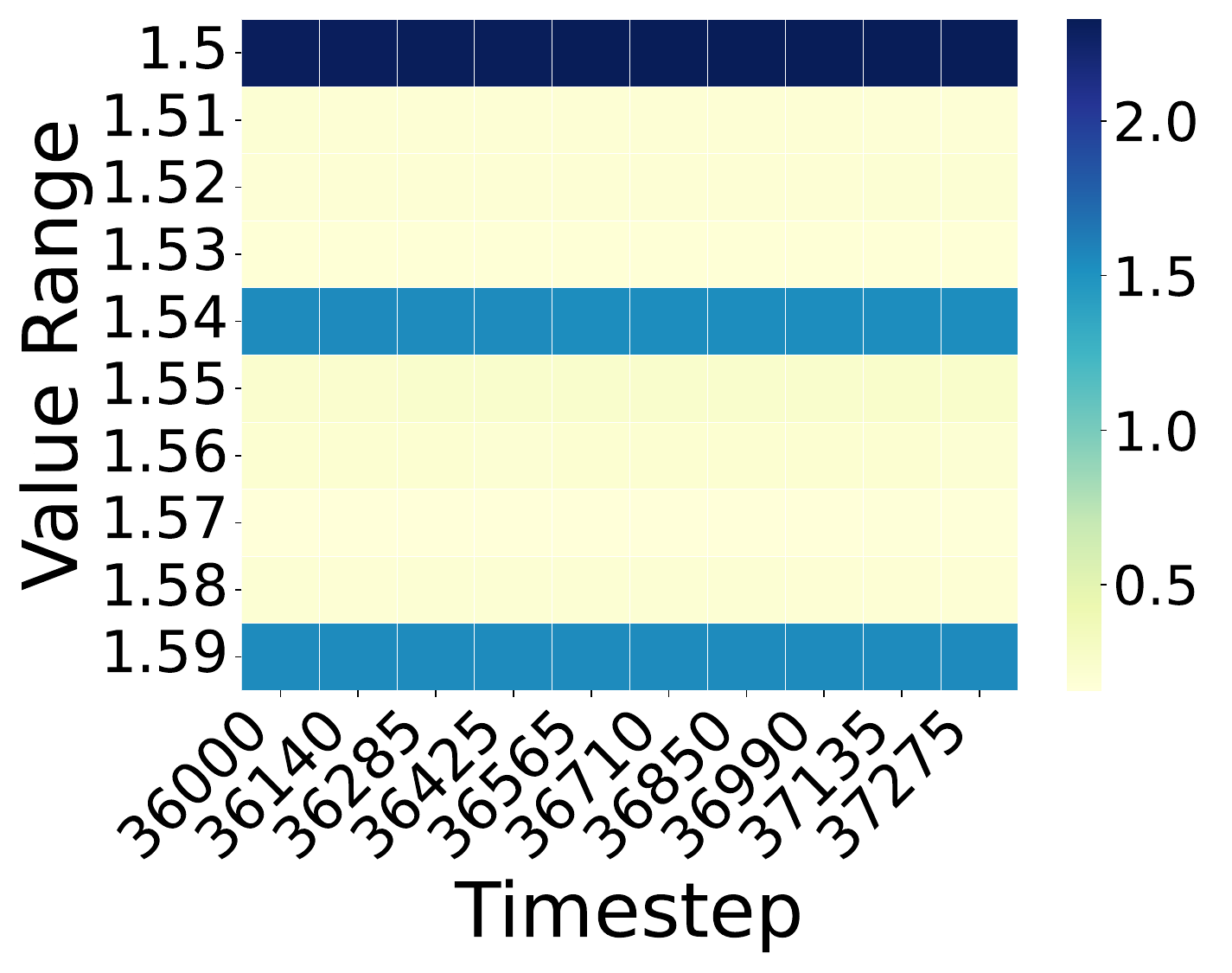} &
    \includegraphics[width=0.333\linewidth]{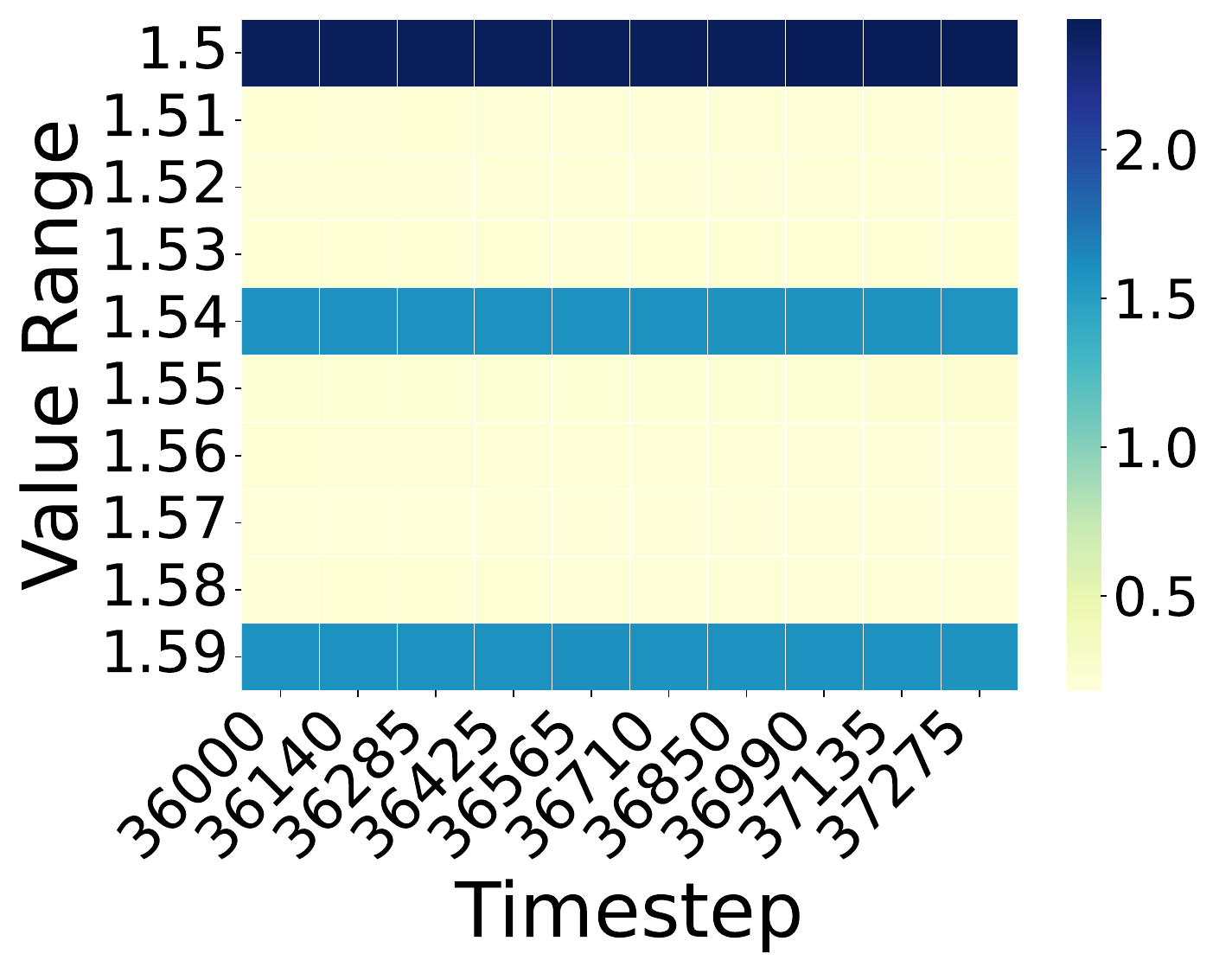} \vspace{-3pt} \\
    \footnotesize (a) Column $x$ &
    \footnotesize (b) Column $y$ &
    \footnotesize (c) Column $z$ \\ 
    \end{tabular}
    \vspace{-5pt}
    \caption{Heatmaps showing distribution rates (‰) of filtered $x$, $y$, and $z$ values across timesteps in the Laghos dataset~\cite{laghos_github}.
    }
    \vspace{-8pt}
    \label{fig:motiv_heatmap}
\end{figure}

\subsection{Low-Selectivity Queries in Scientific Workflows}
\label{sec:low_selectivity}

Prior studies have shown that, in many scientific workflows, queries often require only a small subset of data despite scanning the entire dataset~\cite{Qing_PDSW_2023, Manno_SDC_2023, Zheng_SDC_2022}.  
{To further understand this, we analyzed real-world queries from the CFD domain using the Laghos 3D Mesh Dataset~\cite{lanl_laghos_sample_dataset}} and representative query patterns from Los Alamos National Laboratory (LANL) (refer to the example SQL query in Figure~\ref{fig:back_long}).


Figure~\ref{fig:motiv_heatmap} visualizes the distribution density of records in the Laghos dataset that satisfy the filter condition $1.5 < {x, y, z} < 1.6$. The horizontal axis denotes simulation timestep IDs, and the vertical axis divides the 1.500-1.600 value range into ten uniform bins of width 0.01. Subfigures (a)-(c) respectively illustrate the density distributions for $x$, $y$, and $z$ dimensions (columns).
Most bins exhibit zero density, and even the maximum bin value does not exceed 2‰.  
Given that the query requires all three coordinates to fall within the specified ROI range, only overlapping non-zero bins contribute to the result. The resulting compound selectivity is as low as $1.91\times10^{-4}$\%, indicating an extremely sparse subset of relevant records. This high sparsity reflects a common pattern in domain-specific workflows, where scientists often focus on localized regions or rare events within large-scale simulation outputs~\cite{grider2023snia}.

\subsection{Well-Known Challenges of Data Movement and Storage}
\label{sec:back_challenge}


Scientific workflows produce massive datasets that are persisted to storage and repeatedly moved between storage and compute nodes for post-hoc analysis. 
While distributed engines such as Spark support scalable analysis, their performance is constrained by access to remote data. 
Specifically, modern HPC storage systems face two key limitations that hinder scalable analysis: \textit{data movement} and \textit{data placement}.

\noindent\textbf{Challenge\#1–Massive Data Movement.}
While compute performance in HPC systems has improved rapidly, I/O bandwidth between storage and compute nodes has not scaled accordingly. As a result, data movement from storage to compute nodes dominates analysis runtimes~\cite{grider2023snia}, even when queries access only a small subset of the data. 
This imbalance is exacerbated by growing simulation output sizes and increasing demands for faster turnaround, leading to higher latency, energy consumption, and infrastructure costs~\cite{grider2024snia}.

\vspace{5pt}\noindent\textbf{Challenge\#2–Placement-Access Frequency Mismatch.}
Scientific datasets in columnar formats often exhibit skewed access patterns, where certain columns are accessed far more frequently than others~\cite{IHEP-object,smith2023cephs3objectdata}.
Nevertheless, most HPC storage systems rely on POSIX-based flat file abstractions that lack semantic awareness of column-level locality. This limitation is especially problematic in tiered storage hierarchies comprising HDDs and SSDs, where all data is managed uniformly regardless of access frequency~\cite{IHEP-object}. This mismatch leads to full-file reads from high-latency storage and underutilization of high-performance devices such as NVMe SSDs, degrading bandwidth efficiency and overall system performance.

\subsection{The Advent of Computation-Enabled Object Storage}
\label{sec:back_cos}

\noindent\textbf{Computational Storage for Data Movement Reduction:}
To mitigate the data transfer bottleneck, 
recent efforts have investigated computational storage, which enables basic operations such as \textit{filter} and \textit{project} to be offloaded to the storage layer~\cite{grider2023snia, grider2024snia, KVCSD}. This approach reduces the volume of data transferred to compute nodes, thereby accelerating analysis and alleviating pressure on network and I/O subsystems.
It is particularly effective in low-selectivity scenarios where only a small portion of the dataset satisfies the query predicates.

\noindent\textbf{Improving Data Placement with Object Storage:}~To overcome the limitations of uniform data placement, the HPC community is increasingly adopting object storage~\cite{duwe2021bluestore}.  
Unlike POSIX-based file systems, object storage manages data as discrete objects enriched with metadata, allowing more granular control over storage policies.  
This facilitates adaptive tiering, where frequently accessed columns can be placed on fast NVMe SSDs while rarely accessed data resides on high-capacity HDDs.
Thus, AWS S3~\cite{awss3}-compatible object storage systems such as Ceph~\cite{ceph} and MinIO~\cite{minio} are already in wide use at large-scale research sites such as Conseil Européen pour la Recherche Nucléaire (CERN) and Institute of High Energy Physics (IHEP)~\cite{smith2023cephs3objectdata, riken_fugaku_s3_guide_2024}.
Furthermore, commercial offerings including IBM COS and NetApp StorageGRID which further accelerate this trend~\cite{ibm2024cephs3, netapp2018s3analytics}.  




\noindent{\textbf{Computation-Enabled Object Storage:}}
To address the dual challenges of excessive data movement and suboptimal data placement, recent research has introduced Computation-Enabled Object Storage (COS) systems~\cite{minio-mc-sql, ceph-s3-select, skyhookdm_paper}. 
These systems integrate the flexible, metadata-rich structure of object storage with computation capabilities, enhancing both query efficiency and storage utilization.

There are two main design approaches.
The \textbf{first} extends the object storage interface to support lightweight offloaded computation. 
Systems such as MinIO Select~\cite{minio-mc-sql} and Ceph S3 Select~\cite{ceph-s3-select} enable SQL-like filter queries to be executed directly at the object interface layer, reducing data transfer by pushing down simple operations.
The \textbf{second} approach embeds computation deeper into the storage stack.
A representative example is SkyhookDM~\cite{skyhookdm_paper}, which 
integrates data processing capabilities into Ceph OSD layer using Apache Arrow~\cite{Arrow}.
While SkyhookDM supports basic operations such as \textit{filter} and \textit{aggregate}, it also allows users to define custom processing logic within the storage backend.
By enabling more expressive query offloading at the storage layer, this integration establishes storage-side computation as a foundational building block for scalable and efficient scientific data analysis.

\section{Motivation}
\label{sec:motiv}


While COS solutions have gained attention for supporting query processing at the storage layer, their alignment with the query characteristics of real-world HPC workloads remains largely limited.
The effectiveness of query offloading varies significantly depending on the types of operators involved~\cite{montana2023moveablebeastpartitioningdata, pushdowndb}. 
For instance, \textit{filter} operations, which can substantially reduce data volume, are well-suited for execution at the storage layer. In contrast, high-complexity operations such as \textit{join} incur greater computational overhead and offer limited reductions in data transfer, making them less suitable for offloading.
Nevertheless, there has been little quantitative analysis of the query patterns employed in HPC workloads that process large-scale tabular data, such as those in the CFD, PIC, and HEP domains. Consequently, empirical evidence remains insufficient for evaluating the applicability and limitations of storage-based query offloading in these contexts.

\subsection{Query Characteristics of Tabular HPC Workloads}
\label{sec:observation}
{To quantitatively analyze the query patterns observed in HPC workloads of CFD, HEP, and PIC domains, we collected real-world queries from publicly available analytical workloads released by institutions such as CERN, LANL, and ORNL~\cite{lanl_laghos_sample_dataset, lanl_deep_water_impact_dataset, lanl_c2_vpic_sample_dataset, openPMD-query, canali_spark_hep_2024}}. 
All collected queries operate on tabular scientific datasets. We performed a quantitative structural analysis based on the types of operators used and the ways in which they are composed.

For this analysis, we classified the queries into four categories based on their operator composition:
\squishlist
\item
\textbf{\textit{Filter}}: Queries that include only simple \textit{filter} predicates. 
\item
\textbf{\textit{Filter+Agg/Sort}}: Queries that combine \textit{filter} with \textit{aggregate} or \textit{sort} operations.
\item 
\textbf{\textit{Project}}: Queries that include only \textit{project} operation
\item
\textbf{\textit{Join}}: Queries that involve complex operators such as \textit{join}.
\squishend

\begin{table}[h!]
\centering
\caption{Characteristics of representative queries found in CFD, HEP, and PIC scientific domains.}
\label{tab:query_summary}
\footnotesize
\begin{tabular}{|l|l|c|c|c|c|}
\hline
\textbf{Predicate} & \textbf{Type} & \textbf{\textit{Filter}} & \textbf{\textit{Filter+Agg/Sort}} & \textbf{\textit{Project}} & \textbf{\textit{Join}} \\
\hline\hline
\multirow{2}{*}{Scalar} & Cmp.   & 18 & 2 & 0 & 0 \\
\cline{2-6}
                        & Arith. & 2  & 3 & 9 & 0 \\
\hline
\multirow{2}{*}{Array}  & Cmp.   & 3  & 0 & 0 & 0 \\
\cline{2-6}
                        & Arith. & 10 & 1 & 7 & 0 \\
\hline
\multicolumn{2}{|c|}{User-Defined Func}           & 0 & 0 & 2 & 0 \\
\hline
\multicolumn{2}{|c|}{No Predicate}  & 0 & 0 & 9 & 0 \\
\hline\hline
\multicolumn{2}{|c|}{\textbf{Total}}& 33 & 6 & 27 & 0 \\
\hline
\end{tabular}
\vspace{-4pt}
\end{table}

Table~\ref{tab:query_summary} summarizes the clustering results of queries across each scientific domain. In the table, \textbf{Predicate} indicates whether the data fields used in operator predicates are scalar or array-based, while \textbf{Type} distinguishes between arithmetic (Arith.) and comparison (Comp.) operations.
Among all queries in the table, \bfit{Filter} appeared in 33 cases, \bfit{Filter+Agg/Sort} in 6 cases, and \bfit{Project} in 27 cases. 
Notably, there was not a single instance of a complex query involving a \textit{join} operation across multiple columns. 
The \bfit{Filter} and \bfit{Filter+Agg/Sort} applied predicates exclusively to individual column values, while \bfit{Project} either selected specific columns or generated new ones through column-wise computation. 
These results suggest that the majority of queries operate on a narrow set of columns, without engaging in row-wise processing over entire records. An in-depth analysis further reveals that all queries explicitly reference only the required columns, with non-essential fields excluded from computation.


Publicly available HPC queries primarily consist of simple operations centered around \textit{filter} or \textit{project}. 
This aligns with the typical usage patterns of COS systems and suggests that SQL offloading techniques such as filter pushdown can reduce data transfer and improve analytic performance.
Meanwhile, as shown in the \textbf{Array} section of Table~\ref{tab:query_summary}, complex operator conditions frequently involve computations that go beyond simple scalar comparisons. For example, expressions such as \textit{Muon\_charge[0] != Muon\_charge[1]} involve element-wise comparisons or computations within array-structured columns.
This analysis yields the following three key observations.


\begin{tcolorbox}
\small
\textbf{Observation 1.} Most queries follow a column-based analytics pattern, selectively accessing only the required columns without scanning entire rows.
\\
\textbf{Observation 2.} The core operations are concentrated on \textit{filter}, \textit{project}, \textit{aggregate}, and \textit{sort}, with no occurrence of relational operations such as \textit{join}.
\\
\textbf{Observation 3.} Predicate conditions frequently go beyond simple scalar range comparisons, often involving computations between elements within array-structured columns.
\end{tcolorbox}
These three observations clearly articulate the core requirements for designing next-generation COS storage systems. 
Specifically, such systems should (i) support columnar I/O and processing, (ii) enable efficient in-storage execution of core relational operators such as \textit{filter}, \textit{project}, \textit{aggregate}, and \textit{sort}, and (iii) support predicate evaluation involving element-wise computation over array-valued columns.
Satisfying these requirements is critical to fully realizing the performance advantages of compute offloading in HPC analytical workloads.



\subsection{Limitations of Exisiting COS Systems}
\label{subsec:limit}


\noindent\textbf{Limitation\#1--Limited Output Formats for Columnar Semantics and Compatibility:}
MinIO Select and Ceph S3 Select support filtering at the storage layer but serialize the results in a row-oriented format. This necessitates reconstructing the data into a columnar layout, during which column statistics and structural metadata are lost. As a consequence, downstream analytical engines are unable to apply optimizations such as filter skipping or projection pruning, potentially resulting in redundant reprocessing along the full query path.
SkyhookDM returns query results exclusively in the Arrow IPC format. While this preserves columnar locality, the fixed output format reduces interoperability. Engines that do not provide native support for Arrow, such as  Presto~\cite{presto}, require an additional conversion layer, which complicates integration.
\vspace{5pt}\noindent\textbf{Limitation\#2--Restricted Support for SQL Operators and Array Expressions:}
Current COS solutions exhibit limited operator support and lack the ability to express array-level expressions, even though HPC queries are often simple and structurally well-defined.
MinIO Select and Ceph S3 Select support only simple filtering conditions, basic projection, and regular expression matching, but do not support more advanced operators such as \textit{aggregate}, \textit{sort} or predicate evaluation over array elements~\cite{aws_s3_select_guide}. These limitations prevent the storage layer from processing queries involving array-based predicates and \textit{aggregate} or \textit{sort}, both of which are frequently observed in real HPC workloads. The absence of these features requires entire files or row groups to be transferred to compute nodes, resulting in significant inefficiencies.

SkyhookDM leverages Apache Arrow’s compute kernels to offload \textit{filter} operations, including array element indexing and arithmetic conditions. Although it supports extensibility via user-defined kernels for operations not natively available in Arrow, integrating such kernels requires recompiling the Skyhook-specific libraries within the Arrow ecosystem. This requirement limits flexibility and imposes additional overhead on users.
More critically, \textit{aggregate} and \textit{sort} are not natively supported, and the \textit{project} operator handles only direct column selection. As a result, projections involving computed expressions must be executed on the compute node. Consequently, analytical queries with advanced operators or complex array logic cannot be flexibly offloaded, reducing the overall effectiveness of storage-side computation.


\vspace{5pt}\noindent\textbf{Limitation\#3--Excessive Inter-Layer Data Movement Due to Fixed Execution Layer:}
Existing COS systems typically support computation at only a single logical layer, which leads to structural inefficiencies. For example, systems like MinIO Select or Ceph S3 Select perform operations at the S3 interface level, requiring data to be collected from storage nodes to a gateway node that handles client S3 requests and storage backend coordination. This architecture incurs data movement bottlenecks between storage backend and the gateway.
Even systems such as SkyhookDM, which execute computation at the Object Storage Daemon (OSD) level, are restricted to single-layer execution. This limitation prevents the effective use of emerging lower-tier compute resources, such as DPU-based Just a Bunch of Flash (JBOF) arrays~\cite{supermicrodpu, ddndpu}, whose computational capabilities remain largely underutilized.

In hierarchical storage environments, data movement occurs not only across networks but also between internal layers. Fixing computation at a single layer prevents early-stage reduction and results in excessive inter-layer data transfers. To alleviate this, computations should be initiated at lower layers closer to the data, to progressively reduce volume before reaching upper layers.
\section{Design of \tbdfixed{}}
\label{sec:design}

\subsection{Design Principle}
\label{sec:design_principle}


To address the aforementioned limitations, we design a new COS architecture that preserves the benefits of columnar layout and in-storage computation, while introducing complex operators with array semantics and enabling flexible operator decomposition across execution tiers within the storage stack. Guided by these goals, we present the core Design Principles (DP) that shape the architecture of \tbdfixed{}.

\squishlist
\item
\textbf{DP\#1: Column-Oriented Output Format Support for HPC Analytics.} 
\tbdfixed{} should support both Arrow and CSV outputs to enable efficient columnar processing while preserving compatibility with applications that lack native Arrow support, such as Presto.

\item
\textbf{DP\#2: Storage-Level Query Execution with Advanced Operator and Array Support.} 
As we analyzed in §\ref{sec:observation}, HPC queries often involve \textit{aggregate} and \textit{sort} operations over simulation units, and array-level expressions on physical quantities.  
\tbdfixed{} must support these operators and array semantics for effective and practical in-storage execution.

\item
\textbf{DP\#3: Storage-Aware Query Path Optimization.} 
\tbdfixed{} should reduce internal data movement by generating hierarchical query execution plans that place high data movement cost operations closer to storage and minimize interconnect and network traffic.
\squishend




\subsection{Overview}
\label{sec:design_overview}

The proposed \tbdfixed{} system consists of a \tbdfixed{} frontend (\frontend{}) and multiple storage array (\arraynode{}) servers.
The \frontend{} mainly comprises the following components (§\ref{sec:design_ocsfe}): the \textit{S3 Gateway}, the \textit{Local Optimizer}, the \textit{Metadata Manager}, the \textit{Query Executor and Result Handler}, and the \textit{NVMe-over-Fabrics (NVMe-oF) Initiator module}.
Each \arraynode{} is connected to the \frontend{} via NVMe-oF, enabling high-throughput I/O.  
The \arraynode{} consists of the following components (§\ref{sec:design_ocsa}): the \textit{NVMe-oF Target}, the \textit{Storage Manager}, and the \textit{Query Executor and Result Handler}.

\begin{figure}[!t]
	\centering
	\includegraphics[width=1.0\linewidth]{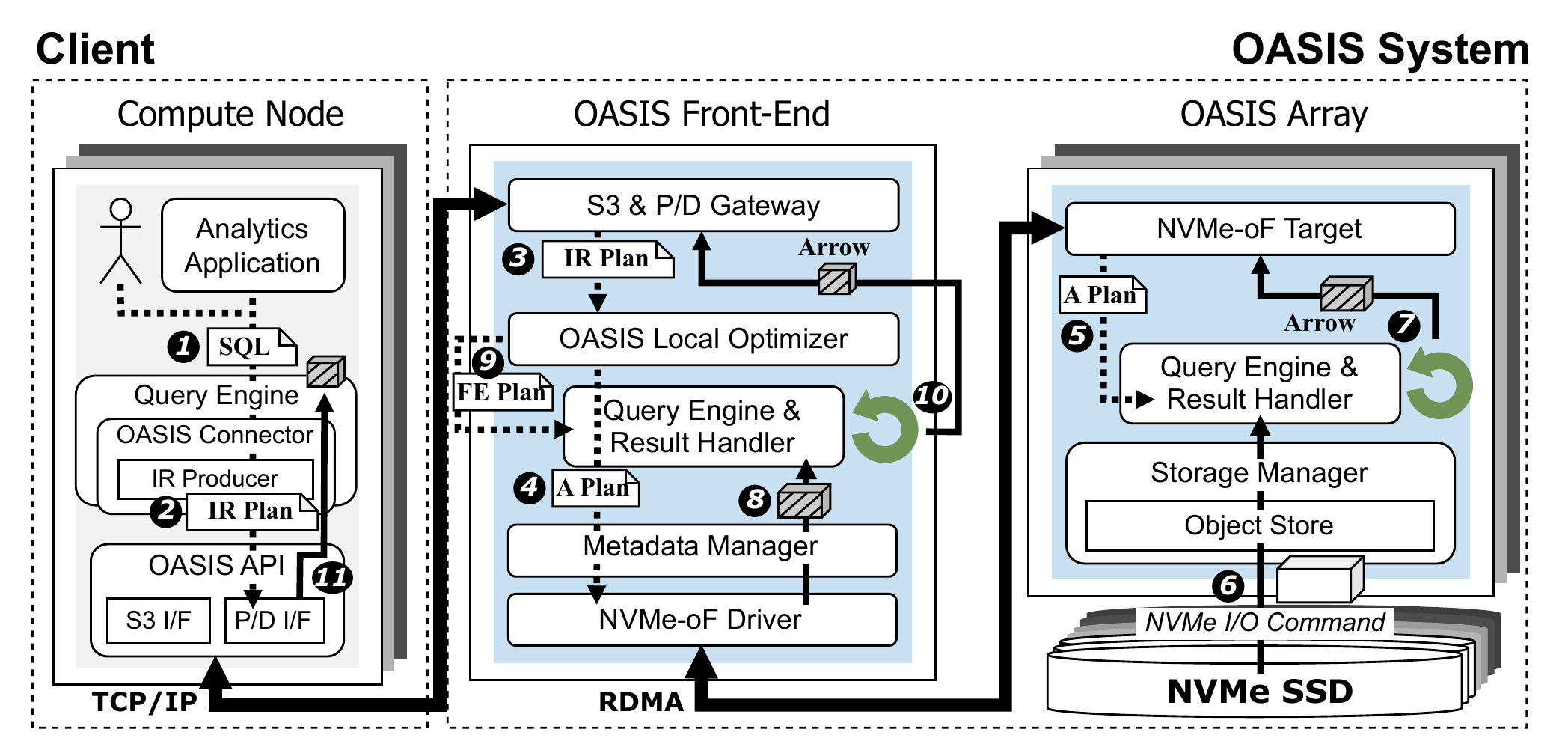}
	\vspace{-15pt}
	\caption{Overall architecture of \tbdfixed{} and a query processing flow. 
    A query task in Substrait IR format is sent to \tbdfixed{} via the P/D Interface (I/F) within  the client-side query engine.}
	\label{fig:architecture}
	\vspace{-10pt}
\end{figure}
\vspace{-3pt}
Figure~\ref{fig:architecture} illustrates the end-to-end query offloading process in the \tbdfixed{} system.
\textbf{\textit{\CircledTextBlack{1}}} The process begins when a client-side query engine submits an SQL query.  
\textbf{\textit{\CircledTextBlack{2}}} The query is translated into an Intermediate Representation (IR), Substrait~\cite{Substrait}, which describes the operator-level execution plan (§\ref{subsec:local_opt}). 
This IR plan is transmitted to the \frontend{} via the \tbdfixed{}-extended Pushdown (P/D) API integrated into the client-side query engine (§\ref{sec:design_client}).  
\textbf{\textit{\CircledTextBlack{3}}} The S3 Gateway forwards the IR plan to the Local Optimizer.  
\textbf{\textit{\CircledTextBlack{4}}} The Local Optimizer partitions the IR plan into a \frontend{} plan and a \arraynode{} plan. 
The \frontend{} plan depends on the intermediate result of the \arraynode{} plan, establishing an execution dependency.  
\textbf{\textit{\CircledTextBlack{5}}} The \arraynode{} plan is sent to the corresponding \arraynode{} for execution.  
\textbf{\textit{\CircledTextBlack{6}}} The target object is retrieved from local storage via the Storage Manager.  
\textbf{\textit{\CircledTextBlack{7}}} The \arraynode{} query engine executes the assigned plan and returns the intermediate result to the \frontend{}.  
\textbf{\textit{\CircledTextBlack{8}}} The \frontend{} registers the result into its query engine as a temporary table.  
\textbf{\textit{\CircledTextBlack{9}}} The \frontend{} executes the \frontend{} plan,  
\textbf{\textit{\CircledTextBlack{10}}} producing and \textbf{\textit{\CircledTextBlack{11}}} returning the final result.


\subsection{Core Components of \frontend{}}
\label{sec:design_ocsfe}

\subsubsection{\textbf{S3 Gateway}}
This module acts as a network-facing entry point that bridges external S3 clients and the internal components of \tbdfixed{}.
It handles S3 protocol parsing and translates standard operations such as \textit{PutObject} and \textit{GetObject} into internal gRPC messages.
These messages encapsulate request metadata and payloads, which are then forwarded to downstream modules for further processing, with responses returned in S3-compatible HTTP format.

\subsubsection{\textbf{Local Optimizer}}
This module analyzes and decomposes offloaded Substrait IR query plans using the Storage-side Query Plan Offloading and Decomposition Algorithm (\vqoa{}).
It estimates the data transfer volume per operator to determine the optimal execution layer, either at the \frontend{} or the \arraynode{}.
Then, the Local Optimizer generates subplans for the \frontend{} and \arraynode{} that account for both operator semantics and communication overhead, enabling efficient query execution while minimizing data movement. 


\subsubsection{\textbf{Metadata Manager}}
This module manages execution-related metadata and performs logical-to-physical translation of requests received from the S3 Gateway.
Upon receiving a gRPC message, it resolves the target object by mapping the S3 bucket name and object key to an internal Object Space ID and Object ID, maintained in its mapping table.
It then generates the corresponding NVMe I/O commands and dispatches them to the appropriate \arraynode{} for execution.
When a new bucket is created, a corresponding \arraynode{} is designated and allocates a unique Object Space ID for routing subsequent I/O requests.
Additional metadata such as object size is included in each request.



To assist the Local Optimizer in operator-level query decomposition, the Metadata Manager collects lightweight statistical metadata during object ingestion.
Specifically, when a \textit{PutObject} operation occurs, a compact histogram is generated via sampling and stored locally on the \frontend{} using the object key as an index.
These histograms typically cover 0.5–5\% of the object 
and capture column-level distributions for various data types such as double and int.
The Local Optimizer later uses this metadata to estimate operator selectivity and output size, guiding plan partitioning and placement.


\subsubsection{\textbf{Query Executor and Result Handler}}
\label{subsubsec:executor}



This module provides a lightweight, columnar-aware execution engine capable of running queries directly at the storage tier.
It supports array-level expressions to efficiently evaluate scientific workloads involving nested or repeated structures.  
After execution, results are serialized into client-specified formats such as Arrow for high-throughput transfer or CSV/JSON for legacy compatibility.
Further implementation details are described in §\ref{subsec:Result_Handler}.

\subsubsection{\textbf{NVMe-oF Initiator}}
This component acts as a network interface that issues NVMe I/O commands to \arraynode{} servers over NVMe-oF.
The \frontend{} converts object storage requests into NVMe commands and transmits them over a low-latency RDMA transport.
Object data and metadata are encapsulated into a single extended NVMe command~\cite{nvmespec}.




\subsection{Core Components of \arraynode{}}
\label{sec:design_ocsa}

\subsubsection{\textbf{NVMe-oF Target}}
The NVMe-oF Target module receives NVMe commands 
from the \frontend{} and performs object I/O operations accordingly.  
Each command encapsulates both object data and metadata, allowing the \arraynode{} to directly access memory and execute read/write operations without separate metadata handling.  
By leveraging an extended block command format, this design enables single-command execution with minimal data copying and supports high-throughput memory-based parallel I/O.

\subsubsection{\textbf{Storage Manager}}
The Storage Manager is responsible for handling storage and retrieval requests issued from the \frontend{} using an internal object store.  
It manages data at the blob level through a Blob Property Table~(BPT), which maps Object Space IDs and Object IDs to physical offsets on the storage device. 
Each blob follows an OPEN–RUN–CLOSE lifecycle during I/O, where DMA buffers and I/O channels are initialized for asynchronous execution.  
Write-Ahead Logging~(WAL) ensures metadata consistency, and a slice-based address space is employed to improve scalability and physical alignment.  
While full implementation details are beyond the scope of this paper, this layer adopts core design principles from conventional object storage systems and provides a robust and scalable backend for upper-layer query execution.

\subsubsection{\textbf{Query Executor and Result Handler}}
This component mirrors the functionality of its counterpart in the \frontend{}, executing assigned sub-plans and returning intermediate results.
To enable fast and efficient communication between the \arraynode{} and the \frontend{}, intermediate results are serialized using the Arrow format and streamed back to the \frontend{}.
Further details are described in §\ref{subsec:Result_Handler}.



\subsection{In-Storage Query Execution Engine and Result Transfer}
\label{subsec:Result_Handler}


The Query Executor and Result Handler are core components of \tbdfixed{}, responsible for realizing \textbf{DP\#1} 
and \textbf{DP\#2} within the storage layer by enabling columnar query processing within the storage layer.  
Accordingly, the execution engine must satisfy three key requirements:  
(1) native support for columnar formats such as Arrow,  
(2) evaluation of expressions on individual elements within array-typed columns, and  
(3) a lightweight execution environment that operates reliably under constrained computational resources.


To fulfill these requirements, we adopt DuckDB~\cite{DuckDB} as the core execution engine within the storage layer.  
DuckDB is an open-source, embedded RDBMS optimized for Online Analytical Processing (OLAP) workloads, offering native support for columnar formats like Parquet and Arrow. 
It supports a wide range of SQL operators--including complex aggregations, sorts, and array-level functions--making it particularly effective for offloading diverse query fragments.
Notably, DuckDB supports both in-memory processing and disk-based execution, allowing it to efficiently process large-scale columnar datasets ranging from tens to hundreds of gigabytes without requiring full in-memory loading.  
This makes it well-suited for lightweight, in-storage execution.





To enable execution across hierarchical layers while preserving the benefits of columnar processing for \textbf{DP\#3}, the Query Executor serializes intermediate results and  final outputs in compressed Arrow format for efficient transfer to the upper layer. For compatibility with legacy tools, final outputs can also be emitted in CSV or JSON format.

Arrow’s columnar layout and zero-copy semantics minimize serialization overhead and enable efficient downstream execution.  
DuckDB natively supports this Arrow-based execution pipeline through Arrow Database Connectivity (ADBC)~\cite{holanda2023adbc}, allowing query execution result tables to be serialized into Arrow and exported to external memory buffers. 
These buffers can then be transferred with minimal overhead, using RDMA for intra-system communication and gRPC for external delivery. 
Result Handlers are deployed at both the \frontend{} and the \arraynode{}, where they work in conjunction with their local query engines to forward intermediate or final result tables to the next processing layer.

\subsection{Operator-Level Query Plan Optimizer and Decomposer}
\label{subsec:local_opt}

The Local Optimizer is responsible for analyzing the Substrait IR-based query plan received from the client and partitioning it across execution layers based on operator characteristics and data transfer cost.
It consists of two core components. First, an optimization algorithm (§\ref{subsec:algo}) identifies the optimal decomposition point for execution. Subsequently, the Substrait Decomposer utilizes this point to partition the original plan into two semantically equivalent subplans, thereby enabling efficient query execution across heterogeneous layers.


\vspace{5pt}
\noindent\textbf{Why Substrait IR?.}
\tbdfixed{} adopts the Substrait IR~\cite{Substrait} to enable fine-grained operator-level partitioning and modular execution across layers. 
Substrait is a cross-platform, language-agnostic IR designed for interoperability between query engines and execution backends.
Compared to SQL, it offers a more structured and expressive form, explicitly encoding operator types, input/output schemas, and expression trees. 
This structure allows the system to efficiently isolate and recompose subplans with minimal transformation overhead~\cite{velox_engine}. 
It also facilitates extensibility through its \texttt{extensionURI} mechanism, enabling the use of user-defined functions and non-standard operators with formal external definitions. 
When the client submits a SQL query, the system translates it into a Substrait IR-based query plan, embedding all necessary metadata such as schema and object references (§~\ref{sec:design_client}).

\begin{figure}[!h]
	\centering
    \vspace{-7pt}
    \includegraphics[width=\linewidth]{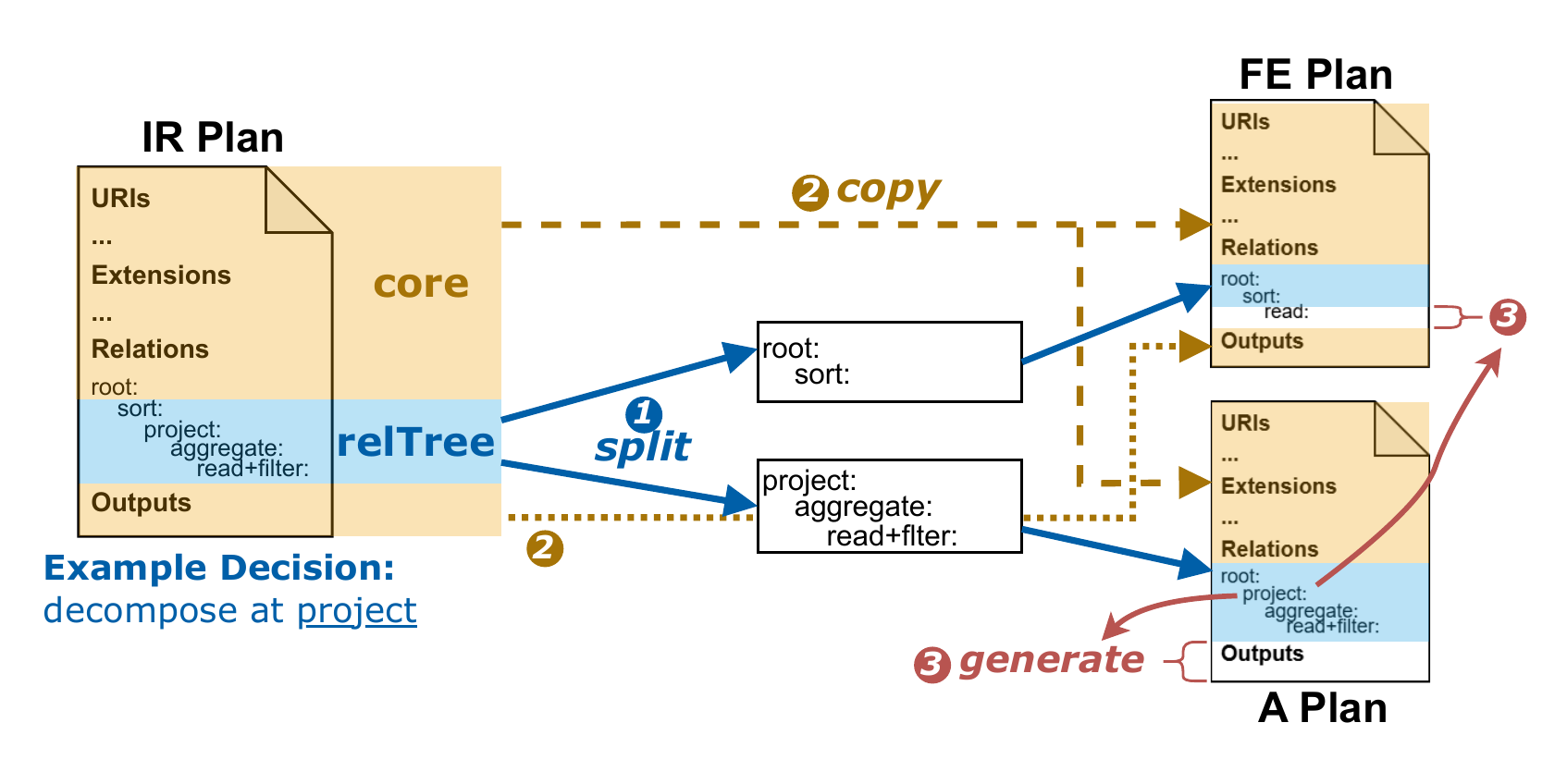}
    \vspace{-20pt}
	\caption{Illustration of the mechanism by which the IR plan is decomposed into the \frontend{} plan and the \arraynode{} plan.
    }
	\label{fig:subs_decomposer}
\end{figure}

\vspace{5pt}
\noindent\textbf{Substrait Decomposer.}
The Substrait Decomposer divides the query plan into two parts based on the operator selected by the optimization algorithm as the decomposition point. 
The decomposition process begins by traversing the original relational tree to locate the target operator that serves as the decomposition point. 
Once identified, the subtree rooted at this operator is extracted to construct the \arraynode{} subplan.
In the original plan, the corresponding node is replaced with a new \textit{read} operator, thereby forming the \frontend{} subplan (Figure~\ref{fig:subs_decomposer}).
\vspace{-8pt}
\vspace{-8pt}
Each resulting subplan is subsequently reconstructed to comply with the Substrait specification.
Specifically, the \texttt{extensionURI} and \texttt{extensions} sections from the original plan are selectively replicated to ensure that all custom operators and functions used in each subplan are correctly defined.
The schema of the intermediate result produced by the \arraynode{} plan is inferred by analyzing the output structure of the extracted subtree, including operators such as \textit{aggregate}, \textit{project}, and \textit{read}. This analysis captures grouping keys, column names, and data types.
To avoid naming conflicts, temporary column names are systematically generated using a unique alphabetical naming convention.
Once the intermediate schema is finalized, it is applied consistently to both subplans. The \arraynode{} plan is explicitly configured to emit this schema, while the \frontend{} plan introduces a new \textit{read} operator that declares the same schema as its input.

This decomposition process forms a semantic bridge between subtree extraction and plan reconstruction. It preserves the logical continuity of data flow while enabling physical modularization of execution. 
Importantly, the resulting plans are not merely structurally separated but also logically ordered: since each operator in the original Substrait IR plan follows a Directed Acyclic Graph (DAG) based on data dependencies, the \arraynode{} plan must be executed first to produce the intermediate output, which is then consumed by the \frontend{} plan. 
This strict dependency preserves the original query semantics while enabling efficient execution across disaggregated compute-storage layers.

\subsection{Optimization Strategies for Query Plan Decomposition}
\label{subsec:algo}

The SODA employed in the Local Optimizer uses two strategies to split Substrait IR plans for HPC tabular queries.
The first strategy, Coefficient-Aware Decomposition (CAD), estimates input-to-output ratios (coefficients) using pre-built histograms and selects a split point that minimizes data movement. It is suited for queries involving scalar-based conditions or simple computations.
The second strategy, Structure-Aware Placement (SAP), applies when coefficients cannot be reliably estimated from statistics or histograms, such as in queries with array-level conditions or computations. SAP considers the physical data layout to place operators close to the data.

\subsubsection{\textbf{Operator Classification}}

It is critical to estimate the data movement introduced by each operator when decomposing a query plan into subplans for the \frontend{} and \arraynode{}. This cost primarily depends on the output size of the last operator executed on the \arraynode{}, since its result is passed to the \frontend{} when execution moves to the upper layer.

If the transition occurs after the $i$-th operator, the output of that operator becomes the input to the $(i{+}1)$-th and constitutes the intermediate data that must be transferred. The size of this intermediate result directly impacts the total data movement overhead.
To quantify this, \tbdfixed{} defines a per-operator input-to-output coefficient based on Substrait operator semantics. Using these coefficients, it estimates the input and output sizes of each operator from the initial input size. Operators are then classified into four categories based on their data transformation behavior, as summarized in Table~\ref{tab:Relation_classification}.

\begin{table}[h]
\centering
\caption{Classification of Substrait operators by type.}
\label{tab:Relation_classification}
\vspace{-5pt}
\setlength{\tabcolsep}{6pt}
\resizebox{\columnwidth}{!}{
{\footnotesize
\begin{tabular}{@{}lcl@{}}
\toprule
\textbf{Type} & \textbf{Input-Output Relationship} & \textbf{Substrait Relations} \\ \midrule
Op 1 & Single parent, $1\!:\!1$ & \textit{read}, \textit{sort} \\
Op 2 & Single parent, $1\!:\!x$ ($x \le 1$) & \textit{filter}, \textit{project}, \textit{aggregate} \\
Op 3 & Single parent, $1\!:\!x$ ($x > 1$) & \textit{expand} \\
Op 4 & Dual parent, $1\!:\!x$ ($x > 0$) & \textit{join}, \textit{set} \\
\bottomrule
\end{tabular}
}
}
\end{table}


As shown in §\ref{sec:observation}, all operators involved in HPC tabular queries fall into either the Op-1 or Op-2 categories. 
Operators in Op-3 and Op-4 (e.g., \textit{join}) are thus excluded from coefficient-based cost estimation. Op-1 operators have identical input and output sizes, yielding fixed coefficients. In contrast, Op-2 operators produce variable output sizes depending on filter selectivity or the number of projected columns, resulting in dynamic coefficients.
To estimate this variability, \tbdfixed{} constructs offline histograms at data ingestion time. These histograms capture column value distributions and are later used to estimate filter selectivity and projection effects, allowing accurate output size estimation for Op-2 operators.

Crucially, coefficient estimation is not only used to predict output sizes but also plays a central role in modeling total data movement. Starting from the input size of the initial \textit{read} operator, \tbdfixed{} performs chained inference across the operator tree, applying estimated coefficients to compute the input and output sizes of subsequent operators.
By combining operator classification, histogram-based estimation, and coefficient inference, \tbdfixed{} builds a cost model focused on data transfer. This enables the system to identify the optimal query plan split point between the \frontend{} and \arraynode{}, minimizing internal data movement and maximizing offloading efficiency.



\subsubsection{\textbf{Coefficient-Aware Decomposition}}





CAD is a strategy designed for query plans involving schemas with scalar-based conditions or computations, where output size can be predicted using coefficient estimation. CAD sequentially infers the input and output sizes of all operators based on their input-output coefficient and the initial input size. 
To determine the optimal split point, we make the following assumption: \textit{Query plans are split under the assumption of one-way data transfer from the lower layer (\arraynode{}) to the upper layer (\frontend{}), without return traffic.}
This assumption avoids unnecessary costs from round-trips and ensures that once sufficient data reduction occurs at the lower layer, intermediate results can be efficiently transferred and processed at the upper layer.






CAD determines the optimal split point through three sequential steps:
\textbf{(1)} Estimate operator-specific input-to-output coefficients using prebuilt histograms;
\textbf{(2)} Propagate input and output size estimates across the operator tree based on these coefficients and the initial input size;
\textbf{(3)} Select a split point based on two criteria: (a) if a semantic boundary requiring centralized processing (e.g., global \textit{sort}) is encountered before further data reduction is possible, the plan is split at that point; (b) If maximal data reduction is achieved, execution continues on the \arraynode{} until a boundary appears, avoiding unnecessary memory transitions and operator materialization in the upper layer.

A representative boundary is the \textit{sort} operator, which requires global ordering and must be merged at the upper layer after partial processing. In contrast, operators like \textit{aggregate} can be safely offloaded, as their commutative and associative properties enable partial aggregation at the lower layer and finalization at the upper layer. Functions like \texttt{MEDIAN}, however, rely on global ordering and cannot be decomposed into partial forms.



\subsubsection{\textbf{Structure-Aware Placement}}




SAP is a decomposition strategy designed for query plans involving array-level conditions or computations, typically found in schemas with nested structures such as \texttt{List} or \texttt{Array}. In such cases, coefficient estimation using Parquet statistics or prebuilt histograms is infeasible, as these are collected at the column level and do not capture intra-array value distributions. Because the output size of such operations depends on runtime evaluations over individual array elements, the CAD strategy is not applicable. 

To address this, SAP mandates that any condition or expression directly referencing array elements be evaluated at the \arraynode{} level. For example, a predicate such as \texttt{a[i] + a[j] < 0}, which depends on the runtime values of individual array elements, cannot be statically predicted and must therefore be executed at the data-resident layer.
SAP proceeds in three steps:
\textbf{(1)} Analyze the query plan to detect array-aware predicates that reference internal items;
\textbf{(2)} Enforce the evaluation of such predicates at the \arraynode{} to ensure locality;
\textbf{(3)} Evaluate the resulting data size at runtime, and apply a lazy execution strategy that transmits results to the \frontend{} only when the output remains within acceptable transfer limits or when the boundary requiring centralized processing is encountered.


By statically determining the plan split while dynamically evaluating result transfer sizes, SAP enables effective offloading even when coefficient-based estimation is infeasible.
This approach supports fine-grained filtering of nested structures near the data, reducing unnecessary transfers and improving overall processing efficiency.

\subsection{Client Integration via IR Producer and Pushdown API}
\label{sec:design_client}


Client-side integration with \tbdfixed{} is enabled via a custom connector, illustrated here using  Spark as a representative example.
The connector consists of two main components (see Figure~\ref{fig:architecture}):
(1) an \textbf{IR Producer} that translates the SQL query into a Substrait IR, and
(2) a \textbf{P/D API} that transmits the IR plan to the \frontend{} via gRPC for pushdown execution.
Users can access data via the standard \texttt{.read.format("...")} interface without modifying their existing Spark applications.
Final query results are serialized in Arrow format and returned to the client, where they can be deserialized directly into Spark DataFrames through the Arrow source interface for further analysis or visualization.
This design enables drop-in compatibility with existing Spark pipelines while leveraging the flexibility of the Substrait IR to support integration with other query engines in the future.

\section{Evaluation}
\label{sec:eval}


\subsection{Experimental Setup}
\label{sec:eval_setup}

\begin{table}[!b]
\centering
\caption{{Details of the hardware specifications used to configure the \tbdfixed{} system and the Spark cluster.}}
\label{tab:eval_setup_combined}
\footnotesize
\resizebox{1.0\columnwidth}{!}{
\begin{tabular}{|l|l|l|}
\hline
\textbf{System} & \textbf{Component} & \textbf{Specification} \\ \hline\hline

\multirow{6}{*}{Spark Cluster} 
& \multirow{3}{*}{Driver} 
    & CPU: Intel\textregistered~Xeon\textregistered~Gold 6226R (3.9~GHz max) \\ \cline{3-3}
& & Cores: 64~cores \\ \cline{3-3}
& & Memory: 386~GB DDR4 \\ \cline{2-3}

& \multirow{3}{*}{Executor} 
    & CPU: Intel\textregistered~Xeon\textregistered~Gold 6330 (3.1~GHz max) \\ \cline{3-3}
& & Cores: 112~cores \\ \cline{3-3}
& & Memory: 128~GB DDR4 \\ \hline\hline

\multirow{7}{*}{\tbdfixed{}} 
& \multirow{3}{*}{\frontend{}} 
    & CPU: Intel\textregistered~Xeon\textregistered~Silver 4410Y (3.9~GHz max) \\ \cline{3-3}
& & Cores: 48~cores \\ \cline{3-3}
& & Memory: 64~GB DDR4 \\ \cline{2-3}

& \multirow{4}{*}{\arraynode{}} 
    & CPU: Intel\textregistered~Xeon\textregistered~Silver 4410Y (2.0~GHz max) \\ \cline{3-3}
& & Cores: 16~cores \\ \cline{3-3}
& & Memory: 64~GB DDR4 \\ \cline{3-3}
& & Storage: 1~TB NVMe SSD + 512~GB SATA SSD \\ \hline

\end{tabular}}
\end{table}

\noindent\textbf{Implementation:}
We implemented a prototype of \tbdfixed{} by building the \arraynode{} using SPDK~\cite{spdk}~v23.09, extending its BDEV layer to incorporate the Storage Manager and Result Handler.
The in-storage Query Executor is built on DuckDB\cite{DuckDB}~v1.3.0.
On the \frontend{}, we employ Versity Gateway\cite{VersityGW}~v2.49.2 for S3 compatibility and implement the Metadata Manager, Result Handler, and Local Optimizer in a C++ backend server.
Communication between the \frontend{} and \arraynode{}s is handled via the NVMe-oF initiator in the Linux kernel~v5.15.0.

\noindent\textbf{\tbdfixed{} and Analytics Cluster Setup:} 
To configure \tbdfixed{}, we used a 48-core, 64~GB memory server as the \frontend{}, and a server with identical specifications but limited to 16~cores as the \arraynode{} and equipped 1~NVMe SSD. 

We deployed a Spark cluster with Spark 4.0.0 consisting of a 64-core, 386\,GB memory server as the Spark driver and two 112-core, 128~GB memory servers as Spark executor nodes.
The server specifications for the \frontend{}, \arraynode{}, and the Spark cluster are listed in Table~\ref{tab:eval_setup_combined}.
The \frontend{} and \arraynode{} are connected via a 10\,GbE RDMA network for SPDK, while \tbdfixed{} communicates with the Spark cluster over 10\,GbE Ethernet network.

\noindent\textbf{Workload:}
For evaluation, we selected three real-world scientific workloads:(1) Laghos~\cite{laghos_github} 3D Mesh Simulation, which models shock hydrodynamics in a Lagrangian framework, (2) DeepWater-Impact Simulation~\cite{deepwater}, which simulates the interaction of water with rigid bodies in marine environments; and (3) CMS Open Data~\cite{CMS-SingleMu-Run2012B-AOD-2013}, which consists of high-energy physics collision event records collected at the CERN LHC.

The Laghos dataset~\cite{lanl_laghos_sample_dataset} is  20GB in size. The DeepWater-Impact workload comprises two datasets~\cite{lanl_deep_water_impact_dataset} of 13GB and 30GB, respectively, which differ in simulation resolution. The CMS dataset contains  12GB of data.
All datasets were converted to the Parquet format for analysis and are publicly available. 

Table~\ref{tab:queries} shows the queries using these datasets. Q1 computes the average energy per vertex within a specified spatial region using the Laghos dataset.
Q2 extracts fluid elements from the Deep Water Impact simulation based on the thresholds.
Q3 analyzes the vertical extent of dynamic fluid activity over time in the high-resolution Deep Water dataset.
Q4 identifies opposite-charge muon pairs with invariant mass between 60 and 120 GeV in CMS event data.


To evaluate the effectiveness of \tbdfixed{} as a COS, we investigate the following research questions~(RQs).

\squishlist
\item \textbf{RQ\#1: } Does \tbdfixed{} provide object-level I/O performance for scientific datasets comparable to existing COSs?
\item \textbf{RQ\#2: }  How effective is \tbdfixed{}’s hierarchical operator execution strategy in improving query performance when evaluated under a uniform storage configuration?
\item \textbf{RQ\#3: } Does the Arrow-based output of \tbdfixed{} offer performance advantages over traditional CSV formats in analytical workflows?
\item \textbf{RQ\#4: } How does selectivity affect the performance of \tbdfixed{} in scientific query workloads?
\item \textbf{RQ\#5: } How effective is \tbdfixed{}'s \vqoa{} strategy compared to other decomposition approaches in hierarchical execution?
\squishend

\noindent\textbf{Comparison:} To answer these research questions, we conducted a series of experiments using the following four configurations.

\squishlist
\item
\textbf{Baseline:} Executes queries using standard Spark processing, where all computation is performed after retrieving data from storage.

\item
\textbf{Pred.:} Extends the Baseline by enabling predicate pushdown to the storage layer.

\item
\textbf{\tbdfixed{}:} Represents our proposed design that employs hierarchical execution across storage layers by \vqoa{}.

\item
\textbf{COS:} Emulates how \tbdfixed{} would operate under the computation model of COSs by executing all operators at the \frontend{}. This configuration ensures that the execution layer follows the same layer as in existing COS systems, allowing to isolate the effects of hierarchical execution without interference such as I/O overhead.
\squishend

\begin{figure}[t]
    \centering
    \begin{minipage}[b]{0.49\linewidth}
        \centering
        \includegraphics[width=\linewidth]{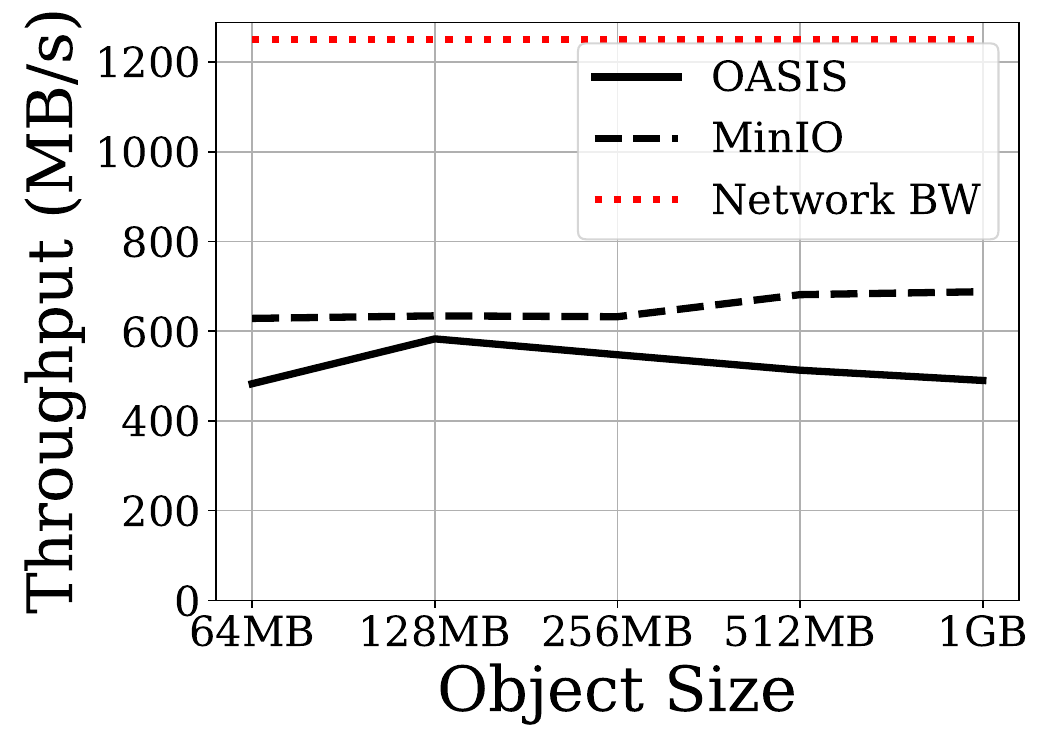}
        \small (a) PUT throughput
    \end{minipage}
    \hfill
    \begin{minipage}[b]{0.49\linewidth}
        \centering
        \includegraphics[width=\linewidth]{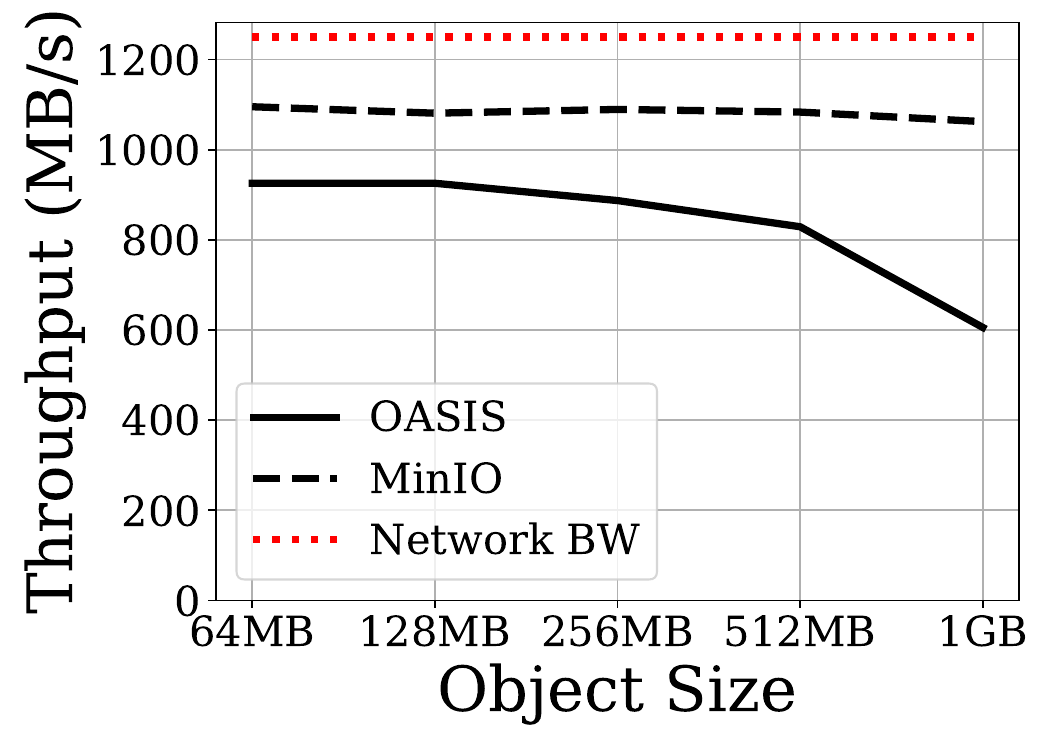}
        \small (b) GET throughput
    \end{minipage}
    \vspace{-10pt}
    \caption{Throughput comparison of OASIS and MinIO.}
    \label{fig:throughput_comparison}
    \vspace{-10pt}
\end{figure}

\subsection{Object-Level I/O Performance}
To evaluate I/O performance for RQ\#1, we compared the object-level PUT and GET throughput of \tbdfixed{} against MinIO, using 16 threads and object sizes from 64MB to 1GB. This reflects typical scientific workloads, where data is written in parallel in 64–256MB blocks, such as Parquet row groups.

Figure~\ref{fig:throughput_comparison} shows the results, with the 10\,Gbps network bandwidth used as the upper bound. In Figure~\ref{fig:throughput_comparison}(a), MinIO achieves up to 681.5MB/s, while \tbdfixed{} peaks at 582.9MB/s but degrades with larger objects due to gRPC overhead from sending many small messages and lack of parallel buffer management. Neither system saturates the 10\,GbE link, mainly due to checksum generation and verification. \tbdfixed{} is further limited by gRPC costs, including message fragmentation and lack of parallel buffer handling. Figure~\ref{fig:throughput_comparison}(b) shows similar trends for GET. MinIO sustains over 1,080MB/s, while \tbdfixed{} drops to 605.6MB/s at 1GB due to the same bottlenecks.
Overall, \tbdfixed{} lags behind MinIO for large objects, but further optimization of buffer management and messaging is expected to close the gap and attain comparable performance.

\begin{table*}[!t]
\centering
\caption{Realistic science discovery queries over datasets from scientific workloads.}
\label{tab:queries}
\vspace{-5pt}
\resizebox{1.9\columnwidth}{!}{%
\scriptsize
\begin{tabular}{ll}
\toprule
\textbf{Query} & \textbf{SQL Statement} \\ \hline
Q1 & 
\begin{tabular}[c]{@{}l@{}}%
SELECT min(vertex\_id) AS VID, min(x) AS X, min(y) AS Y, min(z) AS Z, avg(e) AS E FROM parquet \\%
\quad WHERE x \textgreater{} 1.5 AND x \textless{} 1.6 AND y \textgreater{} 1.5 AND y \textless{} 1.6 AND z \textgreater{} 1.5 AND z \textless{} 1.6 \\%
\quad GROUP BY vertex\_id ORDER BY E;%
\end{tabular} \\ \hline

Q2 & 
\begin{tabular}[c]{@{}l@{}}%
SELECT rowid, v03 FROM parquet \\%
\quad WHERE v03 \textgreater{} 0.001 AND v03 \textless{} 0.999;%
\end{tabular} \\ \hline

Q3 & 
\begin{tabular}[c]{@{}l@{}}%
SELECT MAX((rowid \% (500 * 500)) / 500) AS height, TIMESTEP FROM parquet \\%
\quad WHERE v02 \textgreater{} 0.1 \\%
\quad GROUP BY timestep;%
\end{tabular} \\ \hline

Q4 & 
\begin{tabular}[c]{@{}l@{}}%
SELECT MET\_pt, sqrt( 2 * Muon\_pt[1] * Muon\_pt[2] * (cosh(Muon\_eta[1] - Muon\_eta[2]) - cos(Muon\_phi[1] - Muon\_phi[2]))) \\%
\quad   AS Dimuon\_mass FROM parquet WHERE nMuon = 2
  AND Muon\_charge[1] != Muon\_charge[2] \\ %
  \quad AND sqrt(
           2 * Muon\_pt[1] * Muon\_pt[2] *
           (cosh(Muon\_eta[1] - Muon\_eta[2]) -
            cos(Muon\_phi[1] - Muon\_phi[2]))
      ) BETWEEN 60 AND 120;%
\end{tabular} \\ 
\bottomrule

\end{tabular}
\vspace{-18pt}
}
\end{table*}

\subsection{Effect of Hierarchical Execution on Query Performance}

To evaluate the effectiveness of hierarchical execution in addressing RQ\#2, we performed a series of experiments. \tbdfixed{} leverages \vqoa{} for optimal execution planning, with a detailed analysis provided in §~\ref{subsec:soda}.


\subsubsection{Queries involving Scalar-based Conditions}
\begin{figure}[t]
    \centering
    \includegraphics[width=0.82\linewidth]{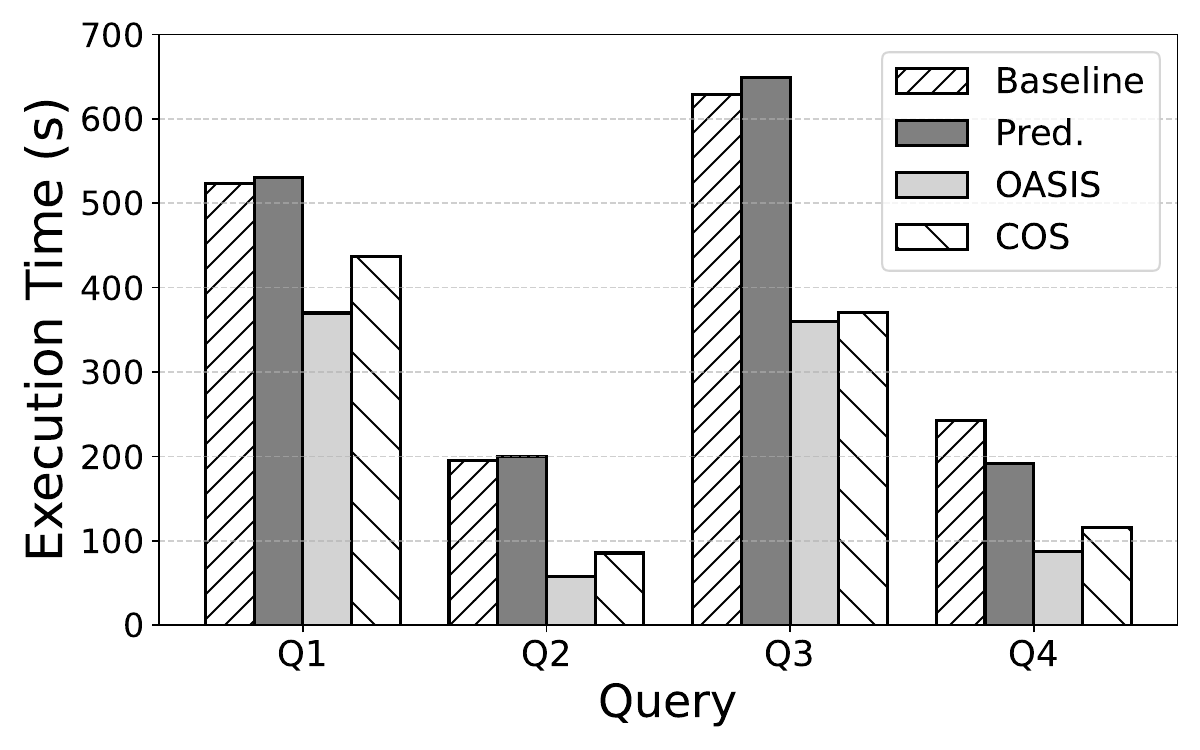}
    \vspace{-5pt}
    \caption{Execution time comparison for the queries across four system configurations.
    Mote that Q1–Q3 use scalar-based conditions, whereas Q4 involves an array-based condition.
    }
    \label{fig:q1_q2_barplot}
    \vspace{-8pt}
\end{figure}

We evaluated three queries (Q1-Q3) involving scalar-based conditions. Each query differs in form, as shown in Table~\ref{tab:queries}. 
As illustrated in Figure~\ref{fig:q1_q2_barplot}, \tbdfixed{} consistently achieves the lowest execution time across both queries. In scalar-based query evaluation, we assume that \textbf{COS} supports all candidate operators, as operator support may vary across systems. This assumption allows us to isolate and validate the impact of hierarchical execution across different queries.

For Q1 and Q2, \textbf{\tbdfixed{}} outperforms \textbf{COS} by 15.27\% in Q1 and 32.7\% in Q2 by minimizing internal data movement between the storage and compute layers, resulting in sharper performance gains. In Q3, \textbf{\tbdfixed{}} continues to deliver the best performance, although the performance gap between \tbdfixed{} and \textbf{COS} narrows. This is because Q3 is compute-intensive, and the performance benefit from reducing data movement becomes less prominent due to the compute capability gap between the \arraynode{} and the \frontend{}.

To validate the results presented in Figure~\ref{fig:q1_q2_barplot}, we measured both the inter-layer data traffic and the size of the result data transferred to the compute layer. 
Across all queries, a consistent trend emerges: both \textbf{COS} and \textbf{\tbdfixed{}} substantially reduce the volume of result data compared to the \textbf{Baseline}. 
For instance, in Q2, the result size is reduced from 13.18GB in the \textbf{Baseline} to 52.89MB (Arrow IPC) in \textbf{\tbdfixed{}} and 29.08~MB (CSV) in \textbf{COS}, with the smaller size in \textbf{COS} attributed to the higher compression ratio of CSV format. 
For inter-layer traffic, \textbf{COS} transfers the entire dataset from the \arraynode{}, while \tbdfixed{} performs early filtering and reduces inter-layer transfer to 53MB. 
This demonstrates that \tbdfixed{} minimizes internal data movement, leading to faster execution for scalar queries with lightweight predicates.
Across all workloads, \textbf{Pred} is slightly slower than the \textbf{Baseline}, primarily due to the overhead of scanning Parquet metadata, with no records being filtered out in the target datasets.

\subsubsection{Queries involving Array-based Conditions}

Queries Q1–Q3 use scalar-based conditions, whereas Q4 involves an array-based condition.
In this section, we evaluated Q4, derived from the HEP benchmark~\cite{Graur_VLDB_2021}, to assess the array-aware processing capabilities of \tbdfixed{}. 
Q4 involves array-based \textit{filter} and \textit{project} operations. 
\textbf{COS}, modeled after vanilla SkyhookDM, supports only array-based \textit{filter} and simple column-level \textit{project}.

Figure~\ref{fig:q1_q2_barplot} illustrates the Q4 execution time of four configurations. \tbdfixed{} completed the query in 87.203 seconds, 24.6\% faster than \textbf{COS} and 64.0\% faster than the \textbf{Baseline}, by offloading both the \textit{filter} and \textit{project} with array-based conditions to the storage layer. 
In contrast, \textbf{COS} offloads only the initial \textit{filter} and must transfer intermediate results to the compute node for array-based \textit{project}, even though it can execute subsequent \textit{filter} operations. This failure to offload \textit{project} results in additional data movement. Predicate Pushdown evaluates only the scalar predicate in storage, showing better performance than \textbf{Baseline}. These results highlight that OASIS’s array-aware offloading reduces data movement and accelerates query execution. 



\begin{figure}[t]
    \centering
    \includegraphics[width=0.82\linewidth]{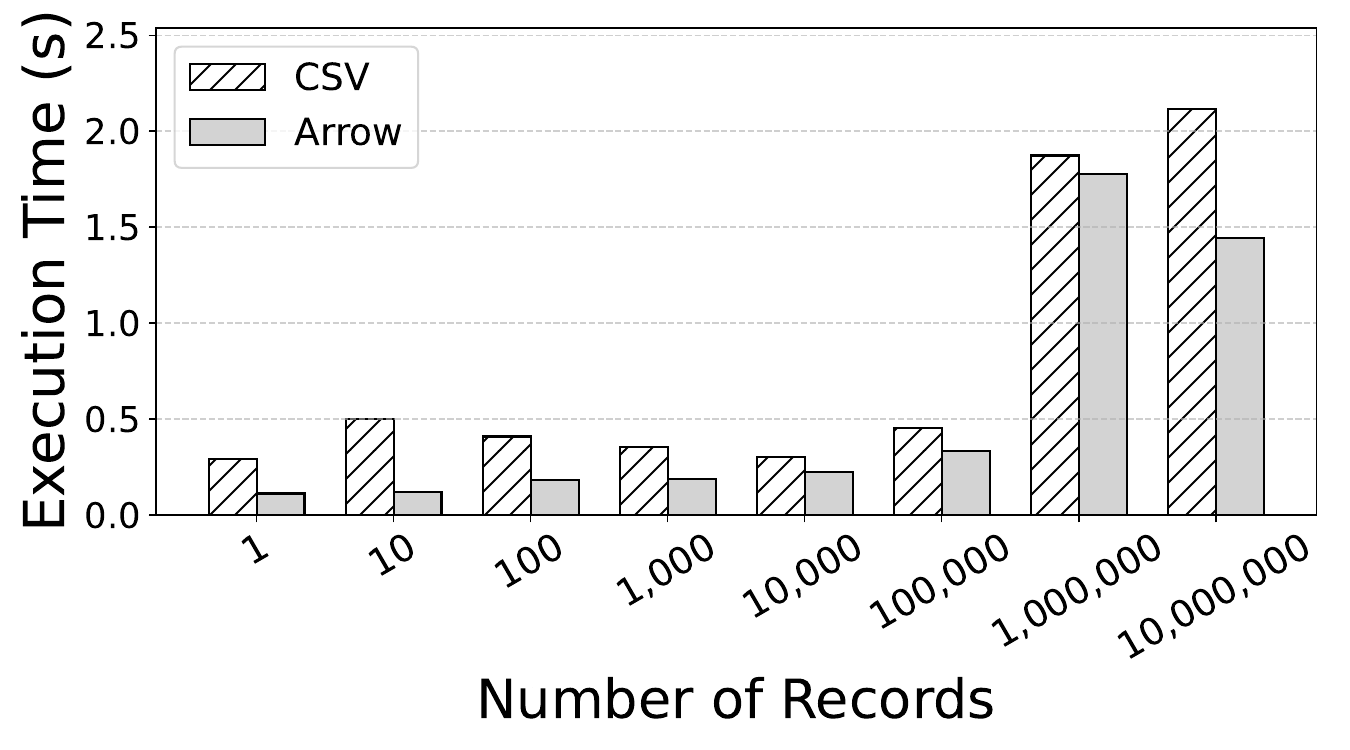}
    \vspace{-5pt}
    \caption{Comparison of input parsing times of client-side Spark for CSV and Arrow formats across different record sizes.
    }
    \label{fig:csv_vs_arrow}
    \vspace{-8pt}
\end{figure}

\subsection{Impact of Output Format on System Efficiency}
\begin{figure*}[t]
    \centering
    \begin{subfigure}[t]{0.40\textwidth}
        \centering
        \includegraphics[width=\linewidth]{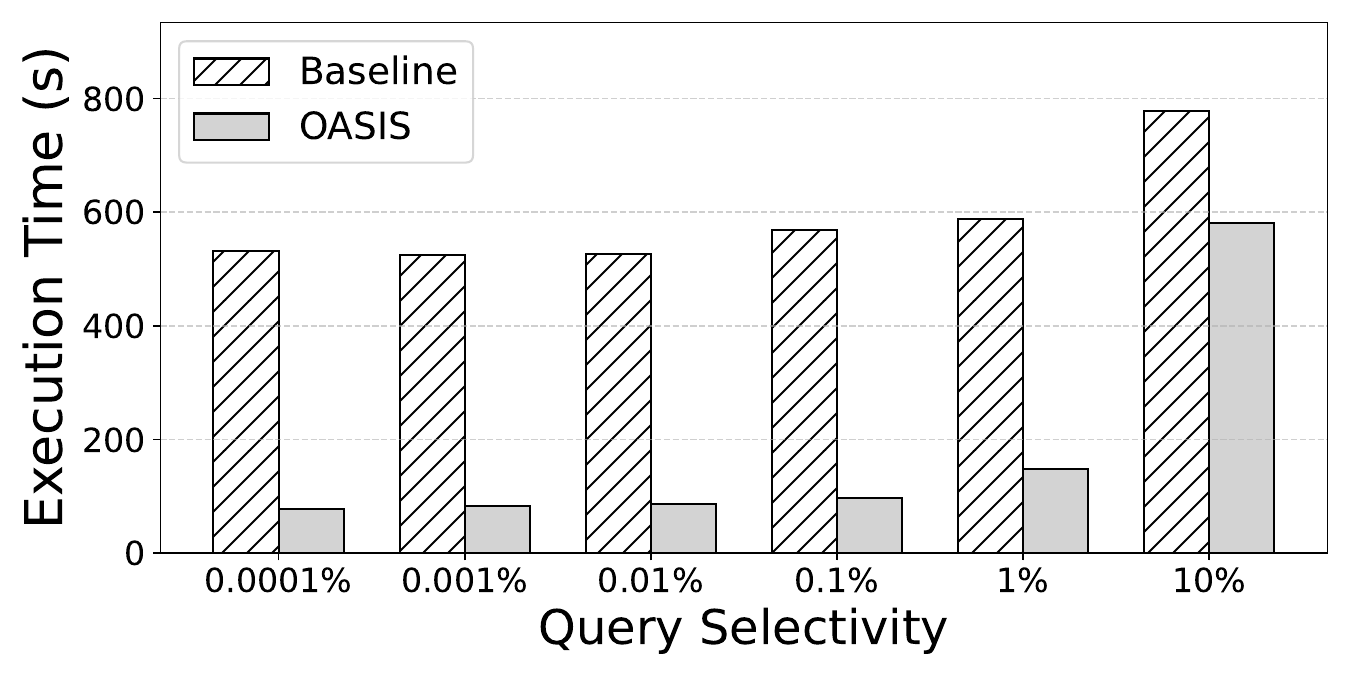}
        \vspace{-18pt}
        \caption{Q1 with aggregation, selectivity: 0.0001\%–10\%.}
        \label{fig:exp_selectivity_groupby}
    \end{subfigure}
    \hspace{15pt}
    \begin{subfigure}[t]{0.40\textwidth}
        \centering
        \includegraphics[width=\linewidth]{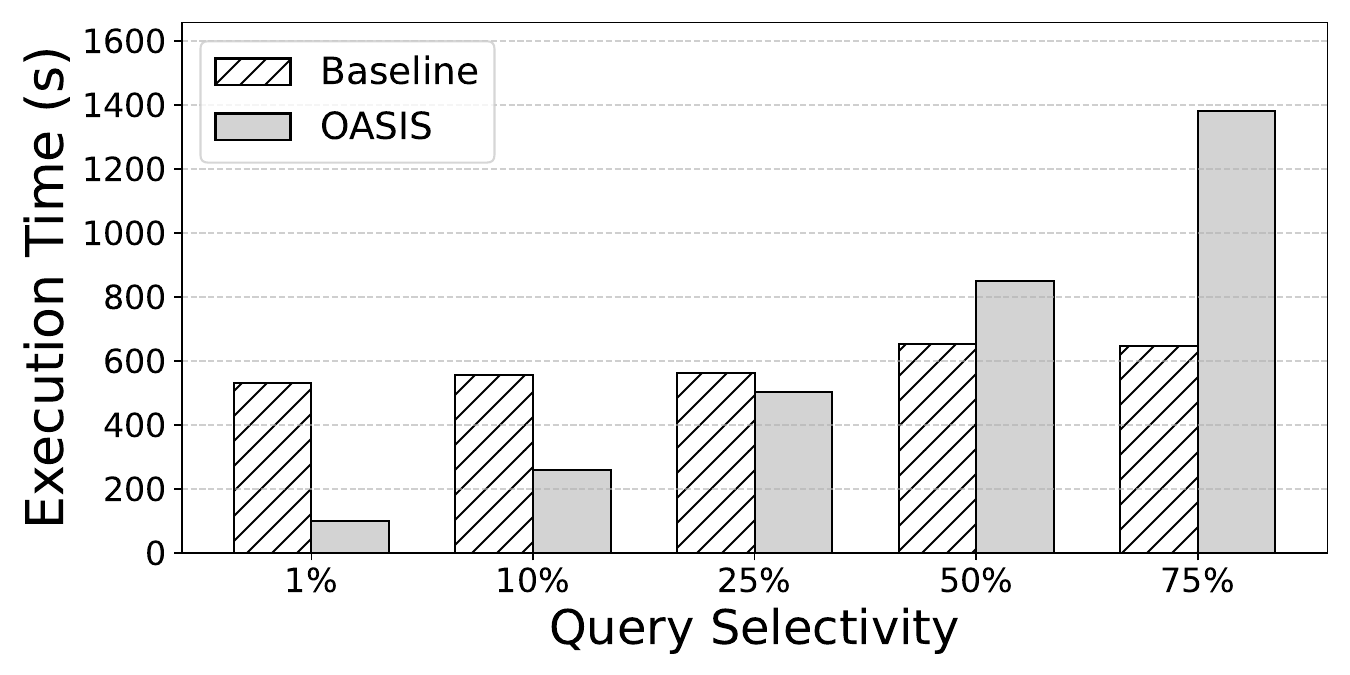}
        \vspace{-18pt}
        \caption{Q1 without aggregation, selectivity: 1\%–75\%.}
        \label{fig:exp_selectivity_noagg}
    \end{subfigure}
    \vspace{-3pt}
    \caption{Execution time comparison between the baseline and \tbdfixed{} for the Q1 query under varying selectivity.
    }
    \label{fig:exp_selectivity_combined}
        \vspace{-8pt}
\end{figure*}

Figure~\ref{fig:csv_vs_arrow} compares the data loading performance of Arrow and CSV formats when ingesting the output of Q1 into the Spark cluster for RQ\#3.
Arrow consistently outperforms CSV across all record sizes in terms of load time.
The observed decrease in execution time from 1,000,000 to 10,000,000 records is due to Spark increasing the number of partitions from 112 to 1,000, thereby enhancing parallelism and improving overall data ingestion throughput.
Since Arrow enables more efficient in-memory loading compared to CSV regardless of data size, these results suggest that Arrow is advantageous not only as a final output format but also as an intermediate data representation during multi-layer query execution.

\subsection{Performance Behavior of \tbdfixed{} Across Diverse Selectivity}

Figure~\ref{fig:exp_selectivity_combined} (a) presents the performance comparison between \tbdfixed{} and the \textbf{Baseline} as the selectivity of Q1 varies. In contrast, (b) shows the results for a modified version of Q1 where the Group By (aggregation) operator is removed, with selectivity similarly adjusted.

In (a), \tbdfixed{} consistently outperforms the \textbf{Baseline} even as selectivity increases. This is because the Group By operation limits the number of output rows based on the number of aggregation groups, preventing the output size from growing rapidly even as input data increases. In fact, for Q1, the maximum achievable selectivity was approximately 13\%, constrained by the nature of the Group By. This suggests that aggregation operations impose a natural upper bound on output size, making them well-suited for query offloading.

On the other hand, (b) explores the case where the Group By operation is removed, allowing selectivity to increase up to approximately 75\%. In this setting, the \textbf{Baseline} begins to outperform \tbdfixed{} when selectivity exceeds around 25\%. This indicates that when heavy operations such as sorting follow the filtering step, traditional cluster-based processing may become more efficient than storage-side offloading as the amount of data grows. These results highlight the need for dynamic offloading decisions based on both the query's operator characteristics and its selectivity.

\begin{figure}[t]
    \centering
    \begin{subfigure}[t]{0.46\linewidth}
        \centering
        \includegraphics[width=\linewidth]{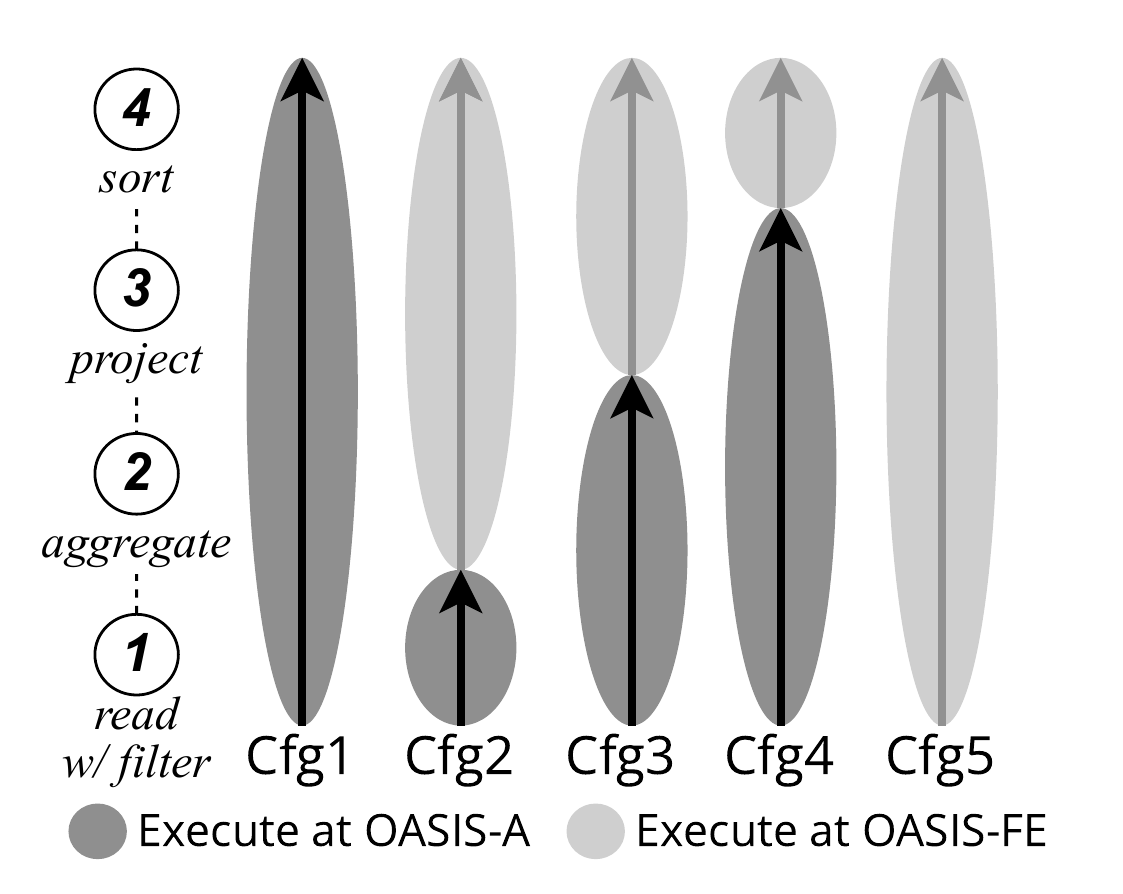}
        \vspace{-18pt}
        \caption{Split configurations}
        \label{fig:cfg_decomposition}
    \end{subfigure}
    \hspace{-2pt}
    \begin{subfigure}[t]{0.5\linewidth}
        \centering
        \includegraphics[width=\linewidth]{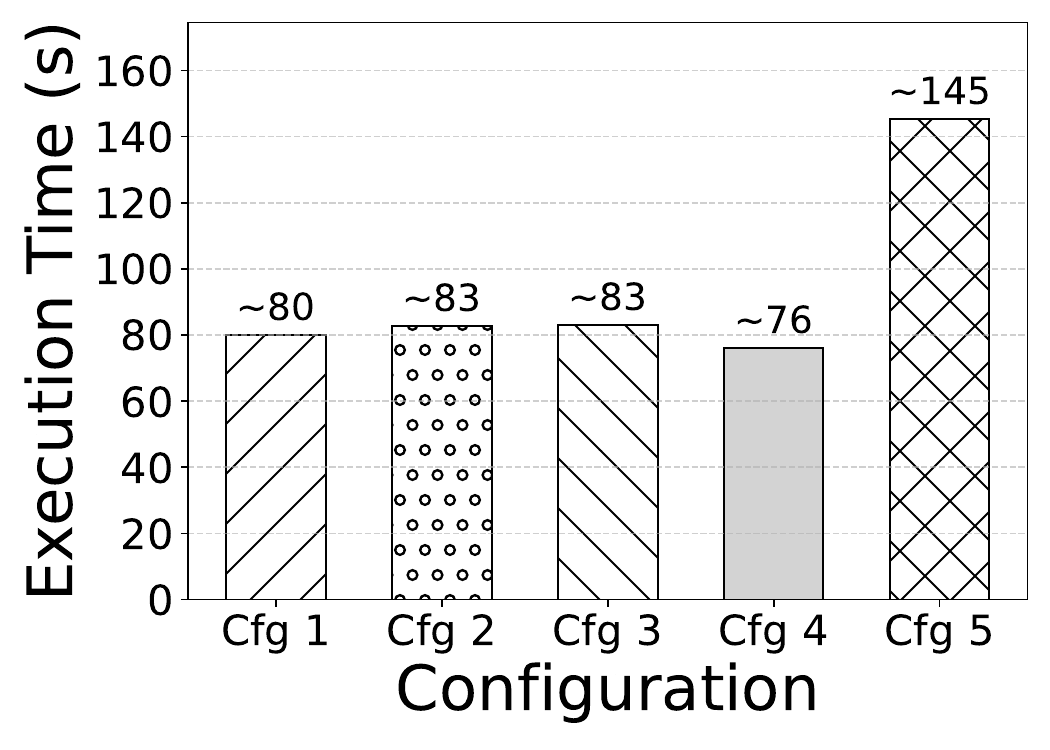}
        \vspace{-18pt}
        \caption{Execution time comparison}
        \label{fig:cfg_execution_time}
    \end{subfigure}
    \vspace{-0pt}
    \caption{(a) Description of decomposition configurations and (b) execution time comparison across five configurations. Each bar uses a distinct hatch pattern for visual distinction.
    }
    \label{fig:cfg_combined}
    \vspace{-8pt}
\end{figure}
\subsection{Effectiveness of \vqoa{} Decomposition}
\label{subsec:soda}
Figure~\ref{fig:cfg_combined} illustrates the evaluation of various split strategies to assess the effectiveness of \vqoa{}.
Among the queries, Q1 contains the largest number of operators, while the others are relatively simple and do not undergo plan splitting.
Therefore, Q1 is selected as the representative case to validate the behavior of \vqoa{}.
Q1’s query plan consists of four sequential stages: \textit{(1) read with filter}, \textit{(2) aggregate}, \textit{(3) project}, and \textit{(4) sort} (Figure~\ref{fig:cfg_combined}(a)).
We evaluate five different configurations within \tbdfixed{}, where the Substrait Decomposer statically distributes operators between the \frontend{} and \arraynode{} without applying \vqoa{}. 
Among all configurations, \vqoa{} selected \texttt{cfg4}, which offloads \textit{read w/ filter}, \textit{aggregate}, and \textit{project} to the \arraynode{}, while executing only \textit{sort} at the \frontend{}. This configuration achieved the best runtime of 76 seconds, yielding a 45\% reduction compared to the \frontend{}-only setup that logically corresponds to the computation model of conventional COS systems. Configurations that offloaded only \textit{filter} or \textit{filter} with \textit{aggregate} showed runtimes around 83 seconds, as the reduced data volume did not fully offset the remaining compute overhead at the \frontend{}. 

In our experiments, \vqoa{} introduces minimal overhead, with an average of just 126ms for selectivity estimation and 1,810ms for Substrait-based plan decomposition.


These results demonstrate that pushing low-cost, high-reduction operators closer to data, while reserving compute-heavy ones like \textit{sort} for the \frontend{}, yields better performance. \vqoa{} can be further improved by incorporating operator-level compute cost into its decision model.

\section{Conclusion}
\label{sec:conc}

In this work, we presented \tbdfixed{}, a computation-enabled object storage (COS) system designed for high-throughput scientific analytics workloads.
\tbdfixed{} overcomes key limitations of existing COS systems by enabling fine-grained operator offloading, supporting complex array-aware expressions, and dynamically optimizing query execution across storage layers.
Leveraging Substrait-based plan decomposition and dynamic execution path optimization, \tbdfixed{} identifies optimal split points to minimize data movement while utilizing in-storage compute.
Real-world HPC query evaluations show that \tbdfixed{} not only reduces execution time but also significantly improves resource efficiency across the storage stack.

\section*{Acknowledgments}
This work was supported by SK hynix Inc. 

\bibliographystyle{ieeetr}
\bibliography{paper}

\end{document}